\documentclass[aps,prl,reprint,superscriptaddress,floatfix,showpacs]{revtex4-1}
\usepackage{xcolor}
\usepackage{amsmath}
\usepackage{amssymb}
\usepackage{empheq}
\usepackage{braket}
\newcommand{\sgn}{\mathop{\mathrm{sgn}}}
\usepackage{bm}
\usepackage{graphicx}
\usepackage[caption=false]{subfig}
\usepackage{hyperref}
\usepackage{siunitx}
\hypersetup{
    colorlinks,
    linkcolor={blue!50!black},
    citecolor={blue!50!black},
    urlcolor={blue!80!black}
}

\begin{document}
\title[Chiral plasmon in gapped Dirac systems]{Chiral plasmon in gapped Dirac systems}
\author{Anshuman Kumar}
\affiliation{Mechanical Engineering Department, Massachusetts Institute of Technology, Cambridge, MA 02139, USA}

\author{Andrei Nemilentsau}
\affiliation{Department of Electrical Engineering \& Computer Science, University of Wisconsin-Milwaukee, Milwaukee, WI 53211, USA}

\author{Kin Hung Fung}
\affiliation{Department of Applied Physics, The Hong Kong Polytechnic University, Hong Kong, China}

\author{George Hanson}
\affiliation{Department of Electrical Engineering \& Computer Science, University of Wisconsin-Milwaukee, Milwaukee, WI 53211, USA}

\author{Nicholas X. Fang}
\email{nicfang@mit.edu}
\affiliation{Mechanical Engineering Department, Massachusetts Institute of Technology, Cambridge, MA 02139, USA}

\author{Tony Low}
\email{tlow@umn.edu}
\affiliation{Department of Electrical \& Computer Engineering, University of Minnesota, Minneapolis, MN 55455, USA}

\date{\today}

\begin{abstract}
We study the electromagnetic response and surface electromagnetic modes in a generic gapped Dirac material under pumping with circularly polarized light. The valley imbalance due to pumping leads to a net Berry curvature, giving rise to a finite transverse conductivity. We discuss the appearance of nonreciprocal chiral edge modes, their hybridization and waveguiding in a nanoribbon geometry, and giant polarization rotation in nanoribbon arrays.
\end{abstract}
\maketitle

\emph{Introduction---} The Berry curvature is a topological property of the Bloch energy band, and acts as an effective  magnetic field in momentum space\cite{berry1984quantal,xiao2010berry,nagaosa2010anomalous}. Hence, topological materials may exhibit anomalous Hall-like transverse currents in the  presence of an applied electric field, in absence of a magnetic field. Examples includes topological insulators\cite{hasan2010colloquium}
with propagating surface states that are protected against backscattering from disorder and impurities, and transition metal dichalcogenides where the two valleys carry opposite Berry curvature giving rise to a bulk topological charge neutral valley currents\cite{xiao2012coupled,mak2014valley}. These bulk topological currents were also experimentally investigated in other Dirac materials, such as gapped graphene and bilayer graphene system\cite{gorbachev2014detecting,sui2015gate}.
The electromagnetic response of these gapped Dirac systems, particularly that due to its surface electromagnetic modes (i.e plasmons) are relative unexplored.

In gapped graphene or transition metal dichalcogenides, electrons in the two valleys have opposite Berry curvature, ensured by time-reversal symmetry (TRS) of their chiral Hamiltonians\cite{xiao2012coupled}. Hence, far field light scattering properties of these system does not differentiate between circularly polarized light, i.e. zero circular dichroism in the classical sense. 
Optical pumping with circularly polarized light naturally breaks TRS, and a net chirality ensues. 
However, under typical experimental conditions, the transverse conductivity due to Berry curvature is less than the quantized conductivity $e^2/\hbar$, and the associated optical dichroism effect is not prominent. These effects, however, can potentially be amplified through enhanced light-matter interaction with plasmons\cite{10.1038/nphoton.2012.262,doi:10.1021/nl201771h,PhysRevB.80.245435,doi:10.1021/nn406627u,Giuliani}.

In this letter, we discuss the emergence of new chiral electromagnetic plasmonic modes and their associated optical dichroism effect. We consider a gapped Dirac system under continuous  pumping with cicularly polarized light. We discuss the appearance of edge chiral plasmons, and how they can allow launching of one-way propagating edge plasmons in a semi-infinite geometry. 
We consider also the hybridization of these chiral edge modes in a  nanoribbon geometry and the possibility of nonreciprocal waveguiding.  Their far-field optical properties reveal resonant absorption accompanied by sizeable polarization rotation.
  \begin{figure}
\includegraphics[width=3.3in]{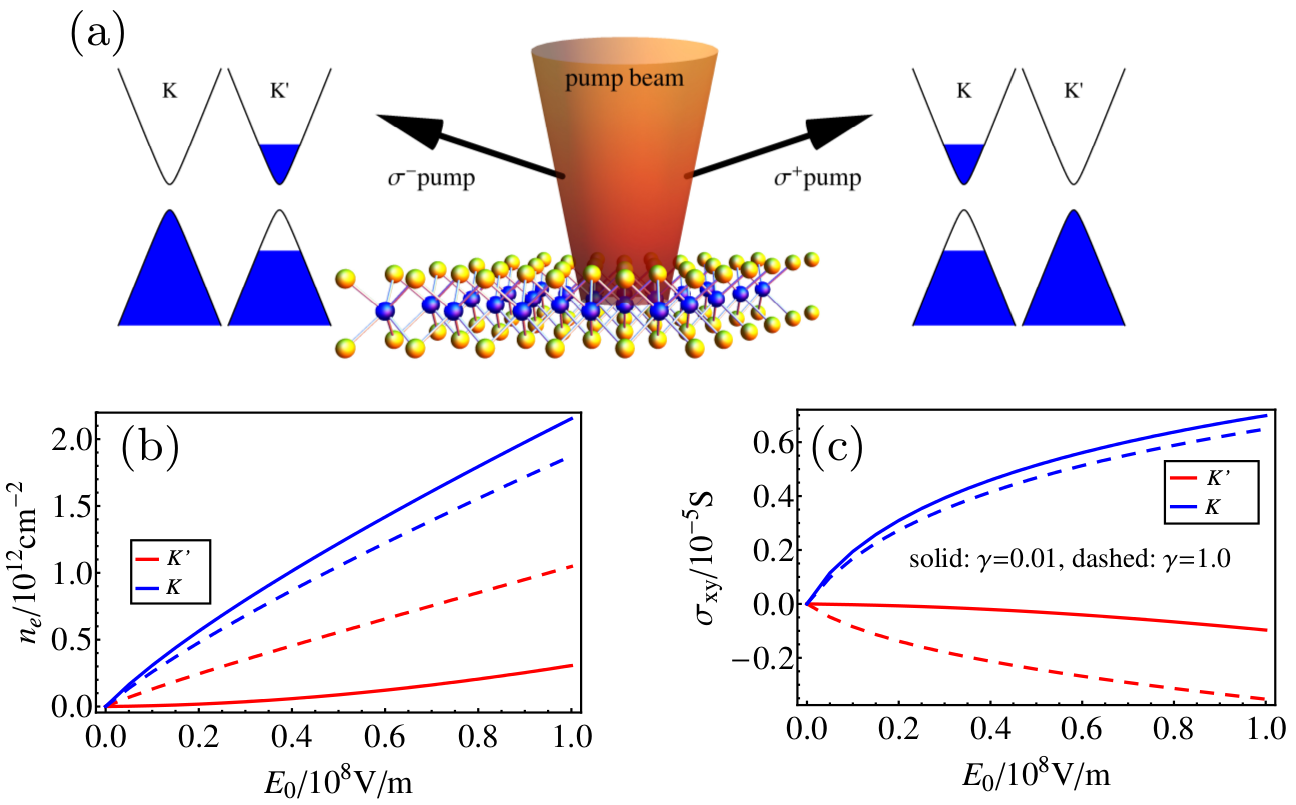}
 \caption{\textit{Optically induced valley polarization}: (a) Polarization selective pumping leads to different populations in the $K$ and $K'$ valleys. (b) DC electronic carrier concentration in the two valleys as a function of the pump electric field.  (c)  DC $\sigma_{xy}$ in the two valleys as a function of the right circular polarized pump electric field}
    \label{fig:MoS2_DC_versus_electric_field}
\end{figure}

\emph{Model system---}
We consider the following Hamiltonian of a massive Dirac system (MDS),
\begin{equation}
{\cal H} = \hbar v_f \mathbf{k}\cdot\boldsymbol{\sigma}_\tau+\tfrac{\Delta}{2}\sigma_z
\label{hamil}
\end{equation}
where $\boldsymbol{\sigma}_\tau =(\tau\sigma_x,\sigma_y) $, $\tau=\pm 1$ denotes the $\bold{K}$/$\bold{K'}$ valley, $\Delta$ is the energy gap and $v_f$ is the Fermi velocity. 
We denote the eigenenergy and wavefunctions of ${\cal H}$ as
${\cal E}_{\tau,\nu}(\bold{k})$ and $\Psi_{\tau,\nu}(\bold{k})$, with $\nu=c,v$ denoting the electron and hole bands. 
We are interested in the dynamics of the electronic subsystem in an external electromagnetic field $\bold{E}$ as illustrated in Fig.\,1a, which can be described with the von Neumann equation, $i\hbar \,\partial_t \hat{\rho}=[ {\cal H} + V, \hat{\rho}  ]$, 
where $\hat{\rho}$ is the statistical operator of the electron subsystem and $V=-e\bold{E}\cdot\bold{r}$ is the interaction term. In the $\Psi_{\tau,\nu}(\bold{k})$ basis, the equation of motion is written explicitly as\cite{Nemilentsau2004,Nemilentsau2006}, 
\begin{eqnarray} \label{Eq:dens_matr}
\nonumber
\frac{\partial \rho_{jj'}}{\partial t}+ \frac{e}{\hbar}\bold{E}\cdot\frac{\partial \rho_{jj'}}{\partial \bold{k}}=-\frac{i}{\hbar}\rho_{jj'}[{\cal E}_{j}(\bold{k}) - {\cal E}_{j'}(\bold{k})]\\
+ \frac{ie}{\hbar}\bold{E}\cdot \sum_{j''}\left[  \bold{R}_{jj''}(\bold{k})  
\rho_{j''j'}-\rho_{jj''}\bold{R}_{j''j'}(\bold{k})
\right]
\end{eqnarray}
where $[\rho(t,\bold{k})]_{j,j'}$ is the density matrix, $\bold{R}_{jj'}(\bold{k})=(i/2)\int
\Psi_{j}^*(\bold{k})\partial_{\mathbf{k}} \Psi_{j'}(\bold{k}) d\bold{r} + hc$, with $j=\{\nu,\tau\}$ designating quantum number of electrons. In Eq. \eqref{Eq:dens_matr} we neglect indirect interband optical transitions.

Here, we are interested in the interaction with a continuous (c.w.) monochromatic electromagnetic wave, $\bold{E}=\bold{E}_0 e^{-i\omega t}+ cc$. Using the rotating wave approximation and introducing relaxation phenomenologically within the relaxation time approximation, we obtain the steady-state solution as a system of four linear equations for diagonal components of the distribution function, $\rho_{j}\equiv\rho_{jj}$,
\begin{eqnarray}
\rho_{\nu,\tau}(\bold{k})=\beta_{\tau}\left(\rho_{\nu,\tau}^{0}(\bold{k})+\gamma \rho_{\nu,\tau'}(\bold{k}) +\alpha_{\tau}\rho_{\nu',\tau}(\bold{k})
\right)
\label{eq:linear_system}
\end{eqnarray}
where  $\tau'\neq\tau$ and $\nu'\neq\nu$. Here,
\begin{eqnarray}
\alpha_{\tau}= \frac{2e^2 |\bold{E}_0\cdot \bold{R}_{cv}^{\tau}(\bold{k})  |^2}{\hbar^2 \left[ (\omega-\omega_{cv})^2+ 1/\tau_0^2   \right]},
\end{eqnarray}
$\omega_{cv}= ({\cal E}_{\tau,c}- {\cal E}_{\tau,v})/\hbar $, $\beta_{\tau}=1/(1+\gamma+\alpha_{\tau})$, $\gamma=\tau_0/\tau_1$, 
where $\tau_0$ is the population relaxation time and $\tau_1$ is the intervalley scattering time (see supplementary information). The equilibrium distribution function is given by the Fermi Dirac distribution, $\rho_{\nu,\tau}^{0}(\bold{k}) = [1+ \mbox{exp}(({\cal E}_{\tau,\nu}(\bold{k})-\mu)/k_B T)]^{-1}  $. 

Let us consider positive or right circular polarized light, $\mathbf{E}_0 = E_0 \left(\mathbf{e}_x + i \mathbf{e}_y\right)$, interacting with electrons at the top of valence band, $\mathbf{R}_{cv}^{\tau}(\mathbf{0}) =  -(v_f\hbar /\Delta)\left(i\tau \mathbf{e}_x + \mathbf{e}_y\right)$. It can be clearly seen that $\bold{E}_0\cdot \bold{R}_{cv}^{\tau}(\mathbf{0}) = -i(v_f\hbar /\Delta) (\tau + 1)$, and thus $\alpha_{\tau}$, are zero at $\mathbf{K}'$ valley while being finite at the $\bold{K}$ valley. Hence, pumping with circularly polarized light would lead to carrier population imbalance between the two valleys.

\emph{Net chirality with pumping---}
The effective Hamiltonian in Eq.\,\ref{hamil} captures the valley physics in physical system such as monolayer graphene with staggered sublattice potential\cite{neto2009electronic} and transition metal dichalcogenides\cite{xiao2012coupled}, if the spin-orbit coupling term can be neglected. To proceed, we consider some reasonable numbers for our model gapped Dirac system; an energy gap $\Delta=0.5$eV and Fermi velocity $v_f=1\times 10^6 \,$ms$^{-1}$. Our calculations assume temperature $T=300\,$K, typical carrier lifetimes $\tau_0=1\,$ps and that the system is undoped at equilibrium. With pumping, charge neutrality and electron-hole symmetry would require that the electron and hole carrier densities follow $n_e^{\tau}= n_h^{\tau} $. Fig.\,1b shows the increasing non-equilibrium electron densities as function of pump intensities $E_0$, under continuous wave  pumping with right circular polarized light. Finite transfers of electrons from the $\bold{K}$ to $\bold{K}'$ valley is determined by the inter-valley scattering rate described by $\gamma$.

In the presence of an external electric field $\bold{E}$, the carrier velocity acquires a non-classical transverse term due to Berry curvature, $\boldsymbol{\Omega}_{\tau}(\bold{k})$, given by $-\tfrac{e}{\hbar}\bold{E}\times\boldsymbol{\Omega}_{\tau}(\bold{k})$. For a MDS, the form of the Berry curvature is well known\cite{xiao2012coupled}.
Within the semiclassical Boltzmann transport theory, 
this would give rise to a transverse conductivity, which in the charge neutral case we are considering here is simply given by
$\sigma^{\tau}_{xy}= 2e^2/\hbar \int [d\bold{k}] \rho_{c,\tau}(\bold{k}) \Omega_{\tau}(\bold{k}) $. The factor of $2$ accounts for contributions from both electrons and holes. It can further be shown that $\sigma^{\tau}_{xy}=-\sigma^{\tau}_{yx}$. Since TRS requires $\Omega_{\bold{K}}(\bold{k}) = - \Omega_{\bold{K}'}(\bold{k})$, $\sigma^{\bold{K}}_{xy} = - \sigma^{\bold{K}'}_{xy}$ at equilibrium. However, under continuous wave pumping, the asymmetric carrier populations in the two valleys would produce a net transverse conductivity, as shown in Fig.\,1c. We note that over the frequency range that we are interested in, i.e. $\hbar\omega \ll \Delta$, $\sigma_{xy}$ is real and frequency independent\cite{PhysRevB.86.205425}. Our calculations suggest that $\sigma_{xy}$ an order smaller than $e^2/\hbar$ is obtainable with pump intensities $E_0$ routinely used in pump-probe experiments. The non-equilibrium longitudinal components of the conductivity, $\sigma_{xx}=\sigma_{yy}$, are computed with the Kubo formula\cite{PhysRevB.86.205425}. For all the subsequent results of this paper, we use a pump intensity of $E_0=10^8$ V/m and $\gamma=0.01$.

\begin{figure}
\includegraphics[width=3.0in]{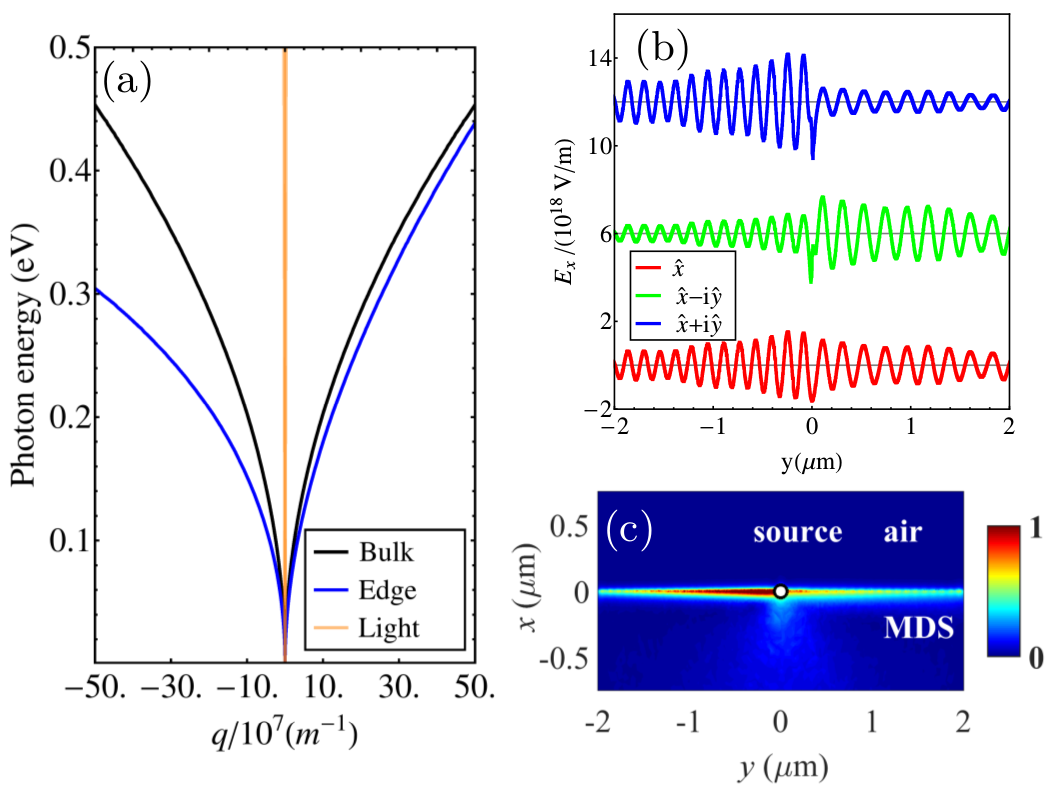}
\caption{\textit{(a) Chiral plasmon dispersion in bulk and semi-infinite MDS.} \textit{(b) Selective excitation of edge modes using circular and linear polarized dipoles placed at the origin:} Line plots of electric field $E_x$ in the plane and perpendicular to the edge. The vertical offset is $6\times10^{18}$V/m. (c) $|E|$ field (normalized to max) profile for a $\hat{x}$-polarized emitter located near the edge of semi infinite Dirac material ($x<0$) at $\omega=\SI{0.1}{eV}$. Both of these field profiles show nonreciprocal emission into the edge mode. The dipoles are placed \SI{10}{nm} above the MDS. }      
\label{fig:MoS2_chiral_plasmon}
\end{figure}

\emph{Valley induced bulk and edge chiral plasmon---}
Armed with the conductivity sum of the two valleys, $\sigma_{ij}$ we discuss general results for the plasmon modes in this system. Plasmon dispersion in a continuous sheet of the MDS is given by\cite{PhysRevB.9.4724,4463896,PhysRevB.91.075419}:
\begin{equation}
\left[ \frac{\epsilon_1}{\kappa_1} + \frac{\epsilon_2}{\kappa_2} + \frac{\imath \sigma_{xx}}{\omega\epsilon_0} \right]\cdot \left[\kappa_1 + \kappa_2 - \frac{\imath\sigma_{yy}}{c\epsilon_0}k_0\right] - \frac{\sigma_{xy}\sigma_{yx}}{(c\epsilon_0)^2}=0
\label{eq:bulk_plasmon}
\end{equation}
where $\kappa_{1,2}=\sqrt{q^2-\epsilon_{1,2}k_0^2}$ are the evanescent decay constants on either side of the 2D sheet. As shown in Fig.~\ref{fig:MoS2_chiral_plasmon}, this ``bulk plasmon" dispersion is symmetric with respect to the wavevector $q$, since it appears quadratically in Eq.~\ref{eq:bulk_plasmon}.

Edges can also accommodate plasmon modes\cite{PhysRevB.32.7676}. Symmetry arguments show that although bulk plasmon dispersion respects $\omega(\mathbf{q})=\omega(-\mathbf{q})$ with nonsymmetric conductivity tensor, the presence of an edge can break this degeneracy\cite{Camley1987103}.  Here we consider the case of semi-infinite MDS. Within the quasistatic picture, the edge plasmon dispersion is approximately given by\cite{PhysRevB.32.7676,PhysRevB.85.235444}: $\eta^2-\chi^2-3\eta + 2\sqrt{2}\chi\sgn(q)=0$, where $\eta=|q|\sigma_{xx}/(\imath\epsilon_0\epsilon\omega)$ and $\chi=|q|\sigma_{xy}/(\epsilon_0\epsilon\omega)$. Fig.~2a indeed shows that the right moving edge plasmon has a different dispersion compared to the left moving one. A simple realization of this nonreciprocity effect consists of placing a dipole near the edge of the material. Finite element simulation of near field dipole emission was performed using \textsc{comsol}. As shown in Fig.~\ref{fig:MoS2_chiral_plasmon}b and c, the linear dipole preferentially emits into the left propagating edge state. Taking a cue from \cite{10.1038/ncomms7695}, we can also use a circular dipole to couple emission into left or right edge state, depending on dipole helicity. The results for circular dipoles are presented in Fig.~\ref{fig:MoS2_chiral_plasmon}b. 

In terms of experiment, the appropriate plasmon momentum can be selected either by use of a grating near the edge\cite{RaetherBook}, by adjusting the distance between the tip of a near field microscope and the edge of the MDS or the tip radius\cite{doi:10.1021/nl202362d,10.1038/nature11253,10.1038/nature11254}. Since the edge plasmon dispersion is nonreciprocal, selection of the magnitude of the plasmon momentum will also lead to selectivity in the propagation direction. In addition to the different intensities of the two, the different wavelengths for the left and right moving edge modes in this configuration might be used for nonreciprocal phase shifters\cite{:/content/aip/journal/apl/99/23/10.1063/1.3665944}. 

\emph{Waveguiding in nanoribbons--} Any practical realization of the semi-infinite case discussed above will involve a stripe or waveguide geometry. Waveguides are an important component of plasmonic circuitry\cite{lsa.2015.67} and ribbon waveguides based on the plasmon modes in graphene have been proposed\cite{PhysRevB.84.161407,doi:10.1021/nn2037626}. In this section, we show how the chirality of the plasmon leads to the propagation direction being coupled to the ribbon edge. It should be noted that unlike the semi-infinite case for ribbons placed in homogeneous space, the dispersion of these plasmon modes is symmetric due to the presence of spatial symmetry\cite{Camley1987103}. However, these modes show nonreciprocity with respect to edge localization as discussed below.

\begin{figure}
\includegraphics[width=2.2in]{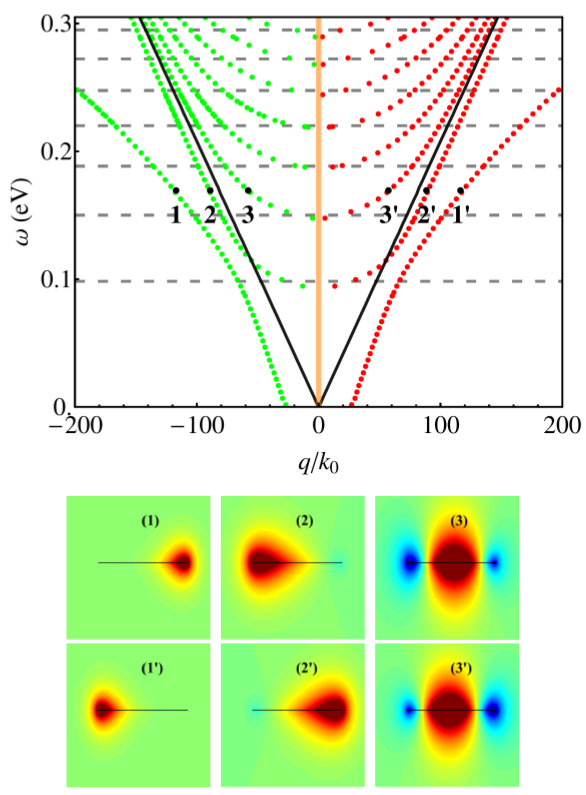}
\caption{\textit{Guided modes in freestanding MDS ribbons.} Ribbon width is assumed to be $w=\SI{100}{nm}$. The grey dashed lines represents solutions of $k_B w + \phi_R=n\pi$, where $k_B$ is the bulk plasmon momentum in MDS and $\phi_R \approx -3\pi/4$\cite{PhysRevB.90.041407}, which explains the cutoff for all the guided modes (except the edge mode). The black solid lines represent the bulk plasmon in a continuous sheet of MDS (same as Fig.~\ref{fig:MoS2_chiral_plasmon}). The color plots below represent the real part of the electric field along the ribbon at the indicated $q$.}
\label{fig:MoS2_ribbon_disp}
\end{figure}

As shown in Fig.~\ref{fig:MoS2_ribbon_disp}, qualitatively the typical profile of plasmons in ribbon\cite{PhysRevB.84.161407,doi:10.1021/nn2037626} or nanowire\cite{doi:10.1021/nl802044t} geometries is observed: there an acoustic branch arising from a monopole like mode and a discrete set of higher order guided modes which show a cutoff. The high frequency field profiles of the two lowest order modes (for example, 1' and 2' for $k_z>0$) in Fig.~\ref{fig:MoS2_ribbon_disp} reveal that these modes have the character of edge modes in semi-infinite MDS. In fact, these edge modes of the ribbon lie outside the ``cone" of the bulk plasmon mode for the continuous MDS. As we approach lower frequencies, the edge localization of these two modes becomes weaker and they start hybridizing.

 All the other modes are guided modes with field maxima in the bulk. These lie inside the cone formed by the dispersion of the plasmon in continuous MDS. Thus these modes are analogous to the guided modes in slab waveguides. The cutoff frequencies for all except the lowest mode are consistent with the Fabry-Per\'ot condition, $k_B w + \phi_R=n\pi$, where $k_B$ is the bulk plasmon momentum in the MDS, as given by Eq.~\ref{eq:bulk_plasmon} and $\phi_R \approx -3\pi/4$ is the approximate phase acquired by the plasmon upon reflection from the ribbon edge\cite{PhysRevB.90.041407}. 
 
The chirality of the plasmon mode in our case, gives rise to the coupling between the propagation direction and the edge. For instance, for positive $k_z$, at higher frequencies, we observe that the field is only confined to the left edge for the lowest mode. Such a coupling is useful for enhancing the lifetime of the mode propagating in a given direction. 

This special coupling between the edge mode direction and the ribbon edge can be utilized to produce explicit nonreciproal devices. For instance, we can break the spatial symmetry between left and right by introducing another medium on one side of the ribbon. In the most extreme case, a perfect conductor can be used to short the edge mode on one side\cite{:/content/aip/journal/apl/99/23/10.1063/1.3665944}.

\emph{Valley induced giant polarization rotation--}
Polarization rotation is usually discussed in the context of magneto-optical materials (also called Faraday effect), where the plane of polarization of the incident wave is rotated upon passage through such a material\cite{Faraday}. Cyclotron resonances in various two dimensional electron gases\cite{doi:10.1143/JPSJ.72.3276} were employed to produce this effect, with graphene being the most promising candidate\cite{10.1038/nphys1816}. Optically induced valley polarization in a MDS presents a promising route to achieve a similar effect without the application of a static magnetic field, which can be cumbersome in the context of on-chip photonic components miniaturization.

We first consider polarization rotation in a continuous sheet of MDS. The polarization rotation angle is given by\cite{PhysRevB.84.235410}: $\theta_F \approx \Re\{\sigma_{xy}\}/2 c\epsilon_0 $ and the transmission by $T(\omega) \approx 1- \Re\{\sigma_{xx}\}/c\epsilon_0$. It should be noted that as opposed to optical activity\cite{10.1038/ncomms1908} which is reciprocal, the polarization rotation in our case is analogous to Faraday rotation which is a purely nonreciprocal effect\cite{0034-4885-67-5-R03}. These equations suggest that the polarization rotation values in the continuous 2D sheet is only dependent on the $\sigma_{xy}$ which can be tuned by adjusting the intensity and polarization of the pump. Even with a pump intensity of the order of ${10^8}${V/m}, rotation angle of only about ${0.1}$ {degrees} is obtained.

\begin{figure}
\includegraphics[width=3.5in]{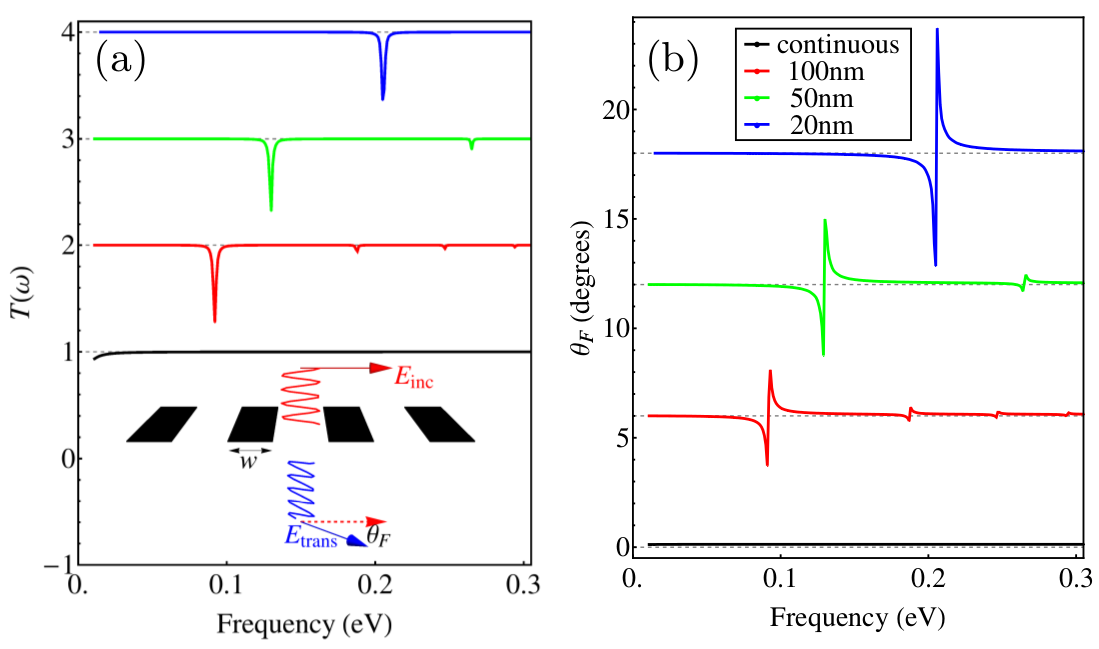}
\caption{\textit{Transmission and polarization rotation in freestanding MDS ribbons}: (a) Transmission (vertical offset is one unit). Inset: Schematic of the configuration. Note that in general the transmitted wave is expected to be elliptically polarized as opposed to linear as shown here. (b) polarization rotation spectrum for different ribbon sizes $w$ (vertical offset is 6 degrees). For ribbon arrays, a filling factor of 50\% has been assumed.}
  \label{fig:mos2_theta}
\end{figure}

However, it is possible to use localized plasmon resonances\cite{PhysRevLett.104.147401,doi:10.1021/ph500278w} of nanoribbons\cite{doi:10.1021/nn403282x} to enhance the polarization rotation values. In Fig.~\ref{fig:mos2_theta}, we present the simulation results for transmission and polarization rotation in nanoribbons. We obtain significant enhancement of polarization rotation by more than an order of magnitude upon using nanoribbons as opposed to a continuous 2D sheet. Moreover, at the frequency of the resonant enhancement, transmitted intensity is still about $10-20 \%$. The spectral location of the resonance is strongly tunable as a function of the ribbon width. These frequencies correspond to the solutions of  $k_B w + \phi_R=n\pi$, as described earlier but with the constraint that $n$ is an even integer\cite{PhysRevB.90.041407}. Odd $n$ solutions are non-dipolar modes, hence do not couple with normally incident plane waves. The largest polarization rotation occurs for smaller ribbon sizes. This is because smaller ribbons, correspond to larger in-plane wavevectors, thus providing a higher field confinement. The polarization rotation we obtained with nanoribbons was found to even surpass Faraday rotation angles in monolayer graphene under a magnetic field of $\SI{7}{T}$\cite{10.1038/nphys1816}. 

\emph{Conclusion and summary---}
In summary, we have shown how polarization selective pumping in a generic gapped Dirac material can impart chirality to bulk and edge plasmons without the need for an external magnetic field. Experimentally testable predictions in the context of near field imaging, giant valley induced polarization rotation as well as nonreciprocal waveguiding were presented. Our theoretical approach can be applied to a general class of two dimensional materials with broken inversion symmetry. A rich array of nonreciprocal phenomenon can be potentially explored, from the point of view of applications to isolators, circulators, etc. Finally, since unlike a magnetic field, the field profile of the optical pump can be easily manipulated on the subwavelength scale by the use of nanostructures\cite{MaierBook}, our work might pave the way for chip scale nonreciprocal photonics and optically tunable metasurfaces\cite{:/content/aip/journal/apl/101/23/10.1063/1.4769095,Kildishev15032013}.

During the preparation of our manuscript, we became aware of a related preprint\cite{arXiv:1506.04743}.

\emph{Acknowledgement.} A.K. and N.X.F. acknowledge the financial support by the NSF (grant CMMI-1120724) and AFOSR MURI (Award No. FA9550-12-1-0488). K.H.F. acknowledges financial support from Hong Kong RGC grant 15300315. T.L. acknowledges support from the MRSEC Program of the National Science Foundation under Award Number DMR-1420013.
\bibliography{Chiral_plasmon_references}

\begin{thebibliography}{47}%
\makeatletter
\providecommand \@ifxundefined [1]{%
 \@ifx{#1\undefined}
}%
\providecommand \@ifnum [1]{%
 \ifnum #1\expandafter \@firstoftwo
 \else \expandafter \@secondoftwo
 \fi
}%
\providecommand \@ifx [1]{%
 \ifx #1\expandafter \@firstoftwo
 \else \expandafter \@secondoftwo
 \fi
}%
\providecommand \natexlab [1]{#1}%
\providecommand \enquote  [1]{``#1''}%
\providecommand \bibnamefont  [1]{#1}%
\providecommand \bibfnamefont [1]{#1}%
\providecommand \citenamefont [1]{#1}%
\providecommand \href@noop [0]{\@secondoftwo}%
\providecommand \href [0]{\begingroup \@sanitize@url \@href}%
\providecommand \@href[1]{\@@startlink{#1}\@@href}%
\providecommand \@@href[1]{\endgroup#1\@@endlink}%
\providecommand \@sanitize@url [0]{\catcode `\\12\catcode `\$12\catcode
  `\&12\catcode `\#12\catcode `\^12\catcode `\_12\catcode `\%12\relax}%
\providecommand \@@startlink[1]{}%
\providecommand \@@endlink[0]{}%
\providecommand \url  [0]{\begingroup\@sanitize@url \@url }%
\providecommand \@url [1]{\endgroup\@href {#1}{\urlprefix }}%
\providecommand \urlprefix  [0]{URL }%
\providecommand \Eprint [0]{\href }%
\providecommand \doibase [0]{http://dx.doi.org/}%
\providecommand \selectlanguage [0]{\@gobble}%
\providecommand \bibinfo  [0]{\@secondoftwo}%
\providecommand \bibfield  [0]{\@secondoftwo}%
\providecommand \translation [1]{[#1]}%
\providecommand \BibitemOpen [0]{}%
\providecommand \bibitemStop [0]{}%
\providecommand \bibitemNoStop [0]{.\EOS\space}%
\providecommand \EOS [0]{\spacefactor3000\relax}%
\providecommand \BibitemShut  [1]{\csname bibitem#1\endcsname}%
\let\auto@bib@innerbib\@empty
\bibitem [{\citenamefont {Berry}(1984)}]{berry1984quantal}%
  \BibitemOpen
  \bibfield  {author} {\bibinfo {author} {\bibfnamefont {M.~V.}\ \bibnamefont
  {Berry}},\ }in\ \href@noop {} {\emph {\bibinfo {booktitle} {Proceedings of
  the Royal Society of London A: Mathematical, Physical and Engineering
  Sciences}}},\ Vol.\ \bibinfo {volume} {392}\ (\bibinfo {organization} {The
  Royal Society},\ \bibinfo {year} {1984})\ pp.\ \bibinfo {pages}
  {45--57}\BibitemShut {NoStop}%
\bibitem [{\citenamefont {Xiao}\ \emph {et~al.}(2010)\citenamefont {Xiao},
  \citenamefont {Chang},\ and\ \citenamefont {Niu}}]{xiao2010berry}%
  \BibitemOpen
  \bibfield  {author} {\bibinfo {author} {\bibfnamefont {D.}~\bibnamefont
  {Xiao}}, \bibinfo {author} {\bibfnamefont {M.-C.}\ \bibnamefont {Chang}}, \
  and\ \bibinfo {author} {\bibfnamefont {Q.}~\bibnamefont {Niu}},\ }\href@noop
  {} {\bibfield  {journal} {\bibinfo  {journal} {Reviews of modern physics}\
  }\textbf {\bibinfo {volume} {82}},\ \bibinfo {pages} {1959} (\bibinfo {year}
  {2010})}\BibitemShut {NoStop}%
\bibitem [{\citenamefont {Nagaosa}\ \emph {et~al.}(2010)\citenamefont
  {Nagaosa}, \citenamefont {Sinova}, \citenamefont {Onoda}, \citenamefont
  {MacDonald},\ and\ \citenamefont {Ong}}]{nagaosa2010anomalous}%
  \BibitemOpen
  \bibfield  {author} {\bibinfo {author} {\bibfnamefont {N.}~\bibnamefont
  {Nagaosa}}, \bibinfo {author} {\bibfnamefont {J.}~\bibnamefont {Sinova}},
  \bibinfo {author} {\bibfnamefont {S.}~\bibnamefont {Onoda}}, \bibinfo
  {author} {\bibfnamefont {A.}~\bibnamefont {MacDonald}}, \ and\ \bibinfo
  {author} {\bibfnamefont {N.}~\bibnamefont {Ong}},\ }\href@noop {} {\bibfield
  {journal} {\bibinfo  {journal} {Reviews of modern physics}\ }\textbf
  {\bibinfo {volume} {82}},\ \bibinfo {pages} {1539} (\bibinfo {year}
  {2010})}\BibitemShut {NoStop}%
\bibitem [{\citenamefont {Hasan}\ and\ \citenamefont
  {Kane}(2010)}]{hasan2010colloquium}%
  \BibitemOpen
  \bibfield  {author} {\bibinfo {author} {\bibfnamefont {M.~Z.}\ \bibnamefont
  {Hasan}}\ and\ \bibinfo {author} {\bibfnamefont {C.~L.}\ \bibnamefont
  {Kane}},\ }\href@noop {} {\bibfield  {journal} {\bibinfo  {journal} {Reviews
  of Modern Physics}\ }\textbf {\bibinfo {volume} {82}},\ \bibinfo {pages}
  {3045} (\bibinfo {year} {2010})}\BibitemShut {NoStop}%
\bibitem [{\citenamefont {Xiao}\ \emph {et~al.}(2012)\citenamefont {Xiao},
  \citenamefont {Liu}, \citenamefont {Feng}, \citenamefont {Xu},\ and\
  \citenamefont {Yao}}]{xiao2012coupled}%
  \BibitemOpen
  \bibfield  {author} {\bibinfo {author} {\bibfnamefont {D.}~\bibnamefont
  {Xiao}}, \bibinfo {author} {\bibfnamefont {G.-B.}\ \bibnamefont {Liu}},
  \bibinfo {author} {\bibfnamefont {W.}~\bibnamefont {Feng}}, \bibinfo {author}
  {\bibfnamefont {X.}~\bibnamefont {Xu}}, \ and\ \bibinfo {author}
  {\bibfnamefont {W.}~\bibnamefont {Yao}},\ }\href@noop {} {\bibfield
  {journal} {\bibinfo  {journal} {Physical Review Letters}\ }\textbf {\bibinfo
  {volume} {108}},\ \bibinfo {pages} {196802} (\bibinfo {year}
  {2012})}\BibitemShut {NoStop}%
\bibitem [{\citenamefont {Mak}\ \emph {et~al.}(2014)\citenamefont {Mak},
  \citenamefont {McGill}, \citenamefont {Park},\ and\ \citenamefont
  {McEuen}}]{mak2014valley}%
  \BibitemOpen
  \bibfield  {author} {\bibinfo {author} {\bibfnamefont {K.~F.}\ \bibnamefont
  {Mak}}, \bibinfo {author} {\bibfnamefont {K.~L.}\ \bibnamefont {McGill}},
  \bibinfo {author} {\bibfnamefont {J.}~\bibnamefont {Park}}, \ and\ \bibinfo
  {author} {\bibfnamefont {P.~L.}\ \bibnamefont {McEuen}},\ }\href@noop {}
  {\bibfield  {journal} {\bibinfo  {journal} {Science}\ }\textbf {\bibinfo
  {volume} {344}},\ \bibinfo {pages} {1489} (\bibinfo {year}
  {2014})}\BibitemShut {NoStop}%
\bibitem [{\citenamefont {Gorbachev}\ \emph {et~al.}(2014)\citenamefont
  {Gorbachev}, \citenamefont {Song}, \citenamefont {Yu}, \citenamefont
  {Kretinin}, \citenamefont {Withers}, \citenamefont {Cao}, \citenamefont
  {Mishchenko}, \citenamefont {Grigorieva}, \citenamefont {Novoselov},
  \citenamefont {Levitov} \emph {et~al.}}]{gorbachev2014detecting}%
  \BibitemOpen
  \bibfield  {author} {\bibinfo {author} {\bibfnamefont {R.}~\bibnamefont
  {Gorbachev}}, \bibinfo {author} {\bibfnamefont {J.}~\bibnamefont {Song}},
  \bibinfo {author} {\bibfnamefont {G.}~\bibnamefont {Yu}}, \bibinfo {author}
  {\bibfnamefont {A.}~\bibnamefont {Kretinin}}, \bibinfo {author}
  {\bibfnamefont {F.}~\bibnamefont {Withers}}, \bibinfo {author} {\bibfnamefont
  {Y.}~\bibnamefont {Cao}}, \bibinfo {author} {\bibfnamefont {A.}~\bibnamefont
  {Mishchenko}}, \bibinfo {author} {\bibfnamefont {I.}~\bibnamefont
  {Grigorieva}}, \bibinfo {author} {\bibfnamefont {K.}~\bibnamefont
  {Novoselov}}, \bibinfo {author} {\bibfnamefont {L.}~\bibnamefont {Levitov}},
  \emph {et~al.},\ }\href@noop {} {\bibfield  {journal} {\bibinfo  {journal}
  {Science}\ }\textbf {\bibinfo {volume} {346}},\ \bibinfo {pages} {448}
  (\bibinfo {year} {2014})}\BibitemShut {NoStop}%
\bibitem [{\citenamefont {Sui}\ \emph {et~al.}(2015)\citenamefont {Sui},
  \citenamefont {Chen}, \citenamefont {Ma}, \citenamefont {Shan}, \citenamefont
  {Tian}, \citenamefont {Watanabe}, \citenamefont {Taniguchi}, \citenamefont
  {Jin}, \citenamefont {Yao}, \citenamefont {Xiao} \emph
  {et~al.}}]{sui2015gate}%
  \BibitemOpen
  \bibfield  {author} {\bibinfo {author} {\bibfnamefont {M.}~\bibnamefont
  {Sui}}, \bibinfo {author} {\bibfnamefont {G.}~\bibnamefont {Chen}}, \bibinfo
  {author} {\bibfnamefont {L.}~\bibnamefont {Ma}}, \bibinfo {author}
  {\bibfnamefont {W.}~\bibnamefont {Shan}}, \bibinfo {author} {\bibfnamefont
  {D.}~\bibnamefont {Tian}}, \bibinfo {author} {\bibfnamefont {K.}~\bibnamefont
  {Watanabe}}, \bibinfo {author} {\bibfnamefont {T.}~\bibnamefont {Taniguchi}},
  \bibinfo {author} {\bibfnamefont {X.}~\bibnamefont {Jin}}, \bibinfo {author}
  {\bibfnamefont {W.}~\bibnamefont {Yao}}, \bibinfo {author} {\bibfnamefont
  {D.}~\bibnamefont {Xiao}},  \emph {et~al.},\ }\href@noop {} {\bibfield
  {journal} {\bibinfo  {journal} {arXiv preprint arXiv:1501.04685}\ } (\bibinfo
  {year} {2015})}\BibitemShut {NoStop}%
\bibitem [{\citenamefont {Grigorenko}\ \emph {et~al.}(2012)\citenamefont
  {Grigorenko}, \citenamefont {Polini},\ and\ \citenamefont
  {Novoselov}}]{10.1038/nphoton.2012.262}%
  \BibitemOpen
  \bibfield  {author} {\bibinfo {author} {\bibfnamefont {A.~N.}\ \bibnamefont
  {Grigorenko}}, \bibinfo {author} {\bibfnamefont {M.}~\bibnamefont {Polini}},
  \ and\ \bibinfo {author} {\bibfnamefont {K.~S.}\ \bibnamefont {Novoselov}},\
  }\href {\doibase 10.1038/nphoton.2012.262} {\bibfield  {journal} {\bibinfo
  {journal} {Nat. Photonics}\ }\textbf {\bibinfo {volume} {6}},\ \bibinfo
  {pages} {749} (\bibinfo {year} {2012})}\BibitemShut {NoStop}%
\bibitem [{\citenamefont {Koppens}\ \emph {et~al.}(2011)\citenamefont
  {Koppens}, \citenamefont {Chang},\ and\ \citenamefont {García~de
  Abajo}}]{doi:10.1021/nl201771h}%
  \BibitemOpen
  \bibfield  {author} {\bibinfo {author} {\bibfnamefont {F.~H.~L.}\
  \bibnamefont {Koppens}}, \bibinfo {author} {\bibfnamefont {D.~E.}\
  \bibnamefont {Chang}}, \ and\ \bibinfo {author} {\bibfnamefont {F.~J.}\
  \bibnamefont {García~de Abajo}},\ }\href {\doibase 10.1021/nl201771h}
  {\bibfield  {journal} {\bibinfo  {journal} {Nano Lett.}\ }\textbf {\bibinfo
  {volume} {11}},\ \bibinfo {pages} {3370} (\bibinfo {year}
  {2011})}\BibitemShut {NoStop}%
\bibitem [{\citenamefont {Jablan}\ \emph {et~al.}(2009)\citenamefont {Jablan},
  \citenamefont {Buljan},\ and\ \citenamefont {Solja\ifmmode \check{c}\else
  \v{c}\fi{}i\ifmmode~\acute{c}\else \'{c}\fi{}}}]{PhysRevB.80.245435}%
  \BibitemOpen
  \bibfield  {author} {\bibinfo {author} {\bibfnamefont {M.}~\bibnamefont
  {Jablan}}, \bibinfo {author} {\bibfnamefont {H.}~\bibnamefont {Buljan}}, \
  and\ \bibinfo {author} {\bibfnamefont {M.}~\bibnamefont {Solja\ifmmode
  \check{c}\else \v{c}\fi{}i\ifmmode~\acute{c}\else \'{c}\fi{}}},\ }\href
  {\doibase 10.1103/PhysRevB.80.245435} {\bibfield  {journal} {\bibinfo
  {journal} {Phys. Rev. B}\ }\textbf {\bibinfo {volume} {80}},\ \bibinfo
  {pages} {245435} (\bibinfo {year} {2009})}\BibitemShut {NoStop}%
\bibitem [{\citenamefont {Low}\ and\ \citenamefont
  {Avouris}(2014)}]{doi:10.1021/nn406627u}%
  \BibitemOpen
  \bibfield  {author} {\bibinfo {author} {\bibfnamefont {T.}~\bibnamefont
  {Low}}\ and\ \bibinfo {author} {\bibfnamefont {P.}~\bibnamefont {Avouris}},\
  }\href {\doibase 10.1021/nn406627u} {\bibfield  {journal} {\bibinfo
  {journal} {ACS Nano}\ }\textbf {\bibinfo {volume} {8}},\ \bibinfo {pages}
  {1086} (\bibinfo {year} {2014})},\ \bibinfo {note} {pMID:
  24484181}\BibitemShut {NoStop}%
\bibitem [{\citenamefont {Giuliani}\ and\ \citenamefont
  {Vignale}(2005)}]{Giuliani}%
  \BibitemOpen
  \bibfield  {author} {\bibinfo {author} {\bibfnamefont {G.}~\bibnamefont
  {Giuliani}}\ and\ \bibinfo {author} {\bibfnamefont {G.}~\bibnamefont
  {Vignale}},\ }\href@noop {} {\emph {\bibinfo {title} {Quantum Theory of the
  Electron Liquid}}}\ (\bibinfo  {publisher} {Cambridge University Press},\
  \bibinfo {year} {2005})\BibitemShut {NoStop}%
\bibitem [{\citenamefont {Slepyan}\ \emph {et~al.}(2004)\citenamefont
  {Slepyan}, \citenamefont {Khrutchinski}, \citenamefont {Nemilentsau},
  \citenamefont {Maksimenko},\ and\ \citenamefont
  {Herrmann}}]{Nemilentsau2004}%
  \BibitemOpen
  \bibfield  {author} {\bibinfo {author} {\bibfnamefont {G.~Y.}\ \bibnamefont
  {Slepyan}}, \bibinfo {author} {\bibfnamefont {A.~A.}\ \bibnamefont
  {Khrutchinski}}, \bibinfo {author} {\bibfnamefont {A.~M.}\ \bibnamefont
  {Nemilentsau}}, \bibinfo {author} {\bibfnamefont {S.~A.}\ \bibnamefont
  {Maksimenko}}, \ and\ \bibinfo {author} {\bibfnamefont {J.}~\bibnamefont
  {Herrmann}},\ }\href {\doibase 10.1142/S0219581X04002152} {\bibfield
  {journal} {\bibinfo  {journal} {International Journal of Nanoscience}\
  }\textbf {\bibinfo {volume} {03}},\ \bibinfo {pages} {343} (\bibinfo {year}
  {2004})}\BibitemShut {NoStop}%
\bibitem [{\citenamefont {Nemilentsau}\ \emph {et~al.}(2006)\citenamefont
  {Nemilentsau}, \citenamefont {Slepyan}, \citenamefont {Khrutchinskii},\ and\
  \citenamefont {Maksimenko}}]{Nemilentsau2006}%
  \BibitemOpen
  \bibfield  {author} {\bibinfo {author} {\bibfnamefont {A.}~\bibnamefont
  {Nemilentsau}}, \bibinfo {author} {\bibfnamefont {G.}~\bibnamefont
  {Slepyan}}, \bibinfo {author} {\bibfnamefont {A.}~\bibnamefont
  {Khrutchinskii}}, \ and\ \bibinfo {author} {\bibfnamefont {S.}~\bibnamefont
  {Maksimenko}},\ }\href {\doibase
  http://dx.doi.org/10.1016/j.carbon.2006.02.035} {\bibfield  {journal}
  {\bibinfo  {journal} {Carbon}\ }\textbf {\bibinfo {volume} {44}},\ \bibinfo
  {pages} {2246 } (\bibinfo {year} {2006})}\BibitemShut {NoStop}%
\bibitem [{\citenamefont {Neto}\ \emph {et~al.}(2009)\citenamefont {Neto},
  \citenamefont {Guinea}, \citenamefont {Peres}, \citenamefont {Novoselov},\
  and\ \citenamefont {Geim}}]{neto2009electronic}%
  \BibitemOpen
  \bibfield  {author} {\bibinfo {author} {\bibfnamefont {A.~C.}\ \bibnamefont
  {Neto}}, \bibinfo {author} {\bibfnamefont {F.}~\bibnamefont {Guinea}},
  \bibinfo {author} {\bibfnamefont {N.}~\bibnamefont {Peres}}, \bibinfo
  {author} {\bibfnamefont {K.~S.}\ \bibnamefont {Novoselov}}, \ and\ \bibinfo
  {author} {\bibfnamefont {A.~K.}\ \bibnamefont {Geim}},\ }\href@noop {}
  {\bibfield  {journal} {\bibinfo  {journal} {Reviews of modern physics}\
  }\textbf {\bibinfo {volume} {81}},\ \bibinfo {pages} {109} (\bibinfo {year}
  {2009})}\BibitemShut {NoStop}%
\bibitem [{\citenamefont {Li}\ and\ \citenamefont
  {Carbotte}(2012)}]{PhysRevB.86.205425}%
  \BibitemOpen
  \bibfield  {author} {\bibinfo {author} {\bibfnamefont {Z.}~\bibnamefont
  {Li}}\ and\ \bibinfo {author} {\bibfnamefont {J.~P.}\ \bibnamefont
  {Carbotte}},\ }\href {\doibase 10.1103/PhysRevB.86.205425} {\bibfield
  {journal} {\bibinfo  {journal} {Phys. Rev. B}\ }\textbf {\bibinfo {volume}
  {86}},\ \bibinfo {pages} {205425} (\bibinfo {year} {2012})}\BibitemShut
  {NoStop}%
\bibitem [{\citenamefont {Chiu}\ and\ \citenamefont
  {Quinn}(1974)}]{PhysRevB.9.4724}%
  \BibitemOpen
  \bibfield  {author} {\bibinfo {author} {\bibfnamefont {K.~W.}\ \bibnamefont
  {Chiu}}\ and\ \bibinfo {author} {\bibfnamefont {J.~J.}\ \bibnamefont
  {Quinn}},\ }\href {\doibase 10.1103/PhysRevB.9.4724} {\bibfield  {journal}
  {\bibinfo  {journal} {Phys. Rev. B}\ }\textbf {\bibinfo {volume} {9}},\
  \bibinfo {pages} {4724} (\bibinfo {year} {1974})}\BibitemShut {NoStop}%
\bibitem [{\citenamefont {Hanson}(2008)}]{4463896}%
  \BibitemOpen
  \bibfield  {author} {\bibinfo {author} {\bibfnamefont {G.}~\bibnamefont
  {Hanson}},\ }\href {\doibase 10.1109/TAP.2008.917005} {\bibfield  {journal}
  {\bibinfo  {journal} {Antennas and Propagation, IEEE Transactions on}\
  }\textbf {\bibinfo {volume} {56}},\ \bibinfo {pages} {747} (\bibinfo {year}
  {2008})}\BibitemShut {NoStop}%
\bibitem [{\citenamefont {Yudin}\ \emph {et~al.}(2015)\citenamefont {Yudin},
  \citenamefont {Eriksson},\ and\ \citenamefont
  {Katsnelson}}]{PhysRevB.91.075419}%
  \BibitemOpen
  \bibfield  {author} {\bibinfo {author} {\bibfnamefont {D.}~\bibnamefont
  {Yudin}}, \bibinfo {author} {\bibfnamefont {O.}~\bibnamefont {Eriksson}}, \
  and\ \bibinfo {author} {\bibfnamefont {M.~I.}\ \bibnamefont {Katsnelson}},\
  }\href {\doibase 10.1103/PhysRevB.91.075419} {\bibfield  {journal} {\bibinfo
  {journal} {Phys. Rev. B}\ }\textbf {\bibinfo {volume} {91}},\ \bibinfo
  {pages} {075419} (\bibinfo {year} {2015})}\BibitemShut {NoStop}%
\bibitem [{\citenamefont {Fetter}(1985)}]{PhysRevB.32.7676}%
  \BibitemOpen
  \bibfield  {author} {\bibinfo {author} {\bibfnamefont {A.~L.}\ \bibnamefont
  {Fetter}},\ }\href {\doibase 10.1103/PhysRevB.32.7676} {\bibfield  {journal}
  {\bibinfo  {journal} {Phys. Rev. B}\ }\textbf {\bibinfo {volume} {32}},\
  \bibinfo {pages} {7676} (\bibinfo {year} {1985})}\BibitemShut {NoStop}%
\bibitem [{\citenamefont {Camley}(1987)}]{Camley1987103}%
  \BibitemOpen
  \bibfield  {author} {\bibinfo {author} {\bibfnamefont {R.}~\bibnamefont
  {Camley}},\ }\href {\doibase http://dx.doi.org/10.1016/0167-5729(87)90006-9}
  {\bibfield  {journal} {\bibinfo  {journal} {Surface Science Reports}\
  }\textbf {\bibinfo {volume} {7}},\ \bibinfo {pages} {103 } (\bibinfo {year}
  {1987})}\BibitemShut {NoStop}%
\bibitem [{\citenamefont {Wang}\ \emph {et~al.}(2012)\citenamefont {Wang},
  \citenamefont {Kinaret},\ and\ \citenamefont {Apell}}]{PhysRevB.85.235444}%
  \BibitemOpen
  \bibfield  {author} {\bibinfo {author} {\bibfnamefont {W.}~\bibnamefont
  {Wang}}, \bibinfo {author} {\bibfnamefont {J.~M.}\ \bibnamefont {Kinaret}}, \
  and\ \bibinfo {author} {\bibfnamefont {S.~P.}\ \bibnamefont {Apell}},\ }\href
  {\doibase 10.1103/PhysRevB.85.235444} {\bibfield  {journal} {\bibinfo
  {journal} {Phys. Rev. B}\ }\textbf {\bibinfo {volume} {85}},\ \bibinfo
  {pages} {235444} (\bibinfo {year} {2012})}\BibitemShut {NoStop}%
\bibitem [{\citenamefont {le~Feber}\ \emph {et~al.}(2015)\citenamefont
  {le~Feber}, \citenamefont {Rotenberg},\ and\ \citenamefont
  {Kuipers}}]{10.1038/ncomms7695}%
  \BibitemOpen
  \bibfield  {author} {\bibinfo {author} {\bibfnamefont {B.}~\bibnamefont
  {le~Feber}}, \bibinfo {author} {\bibfnamefont {N.}~\bibnamefont {Rotenberg}},
  \ and\ \bibinfo {author} {\bibfnamefont {L.}~\bibnamefont {Kuipers}},\ }\href
  {http://dx.doi.org/10.1038/ncomms7695} {\bibfield  {journal} {\bibinfo
  {journal} {Nat Commun}\ }\textbf {\bibinfo {volume} {6}},\  (\bibinfo {year}
  {2015})}\BibitemShut {NoStop}%
\bibitem [{\citenamefont {Raether}(1988)}]{RaetherBook}%
  \BibitemOpen
  \bibfield  {author} {\bibinfo {author} {\bibfnamefont {H.}~\bibnamefont
  {Raether}},\ }\href@noop {} {\emph {\bibinfo {title} {Surface Plasmons on
  Smooth and Rough Surfaces and on Gratings}}}\ (\bibinfo  {publisher}
  {Springer},\ \bibinfo {year} {1988})\BibitemShut {NoStop}%
\bibitem [{\citenamefont {Fei}\ \emph {et~al.}(2011)\citenamefont {Fei},
  \citenamefont {Andreev}, \citenamefont {Bao}, \citenamefont {Zhang},
  \citenamefont {S.~McLeod}, \citenamefont {Wang}, \citenamefont {Stewart},
  \citenamefont {Zhao}, \citenamefont {Dominguez}, \citenamefont {Thiemens},
  \citenamefont {Fogler}, \citenamefont {Tauber}, \citenamefont {Castro-Neto},
  \citenamefont {Lau}, \citenamefont {Keilmann},\ and\ \citenamefont
  {Basov}}]{doi:10.1021/nl202362d}%
  \BibitemOpen
  \bibfield  {author} {\bibinfo {author} {\bibfnamefont {Z.}~\bibnamefont
  {Fei}}, \bibinfo {author} {\bibfnamefont {G.~O.}\ \bibnamefont {Andreev}},
  \bibinfo {author} {\bibfnamefont {W.}~\bibnamefont {Bao}}, \bibinfo {author}
  {\bibfnamefont {L.~M.}\ \bibnamefont {Zhang}}, \bibinfo {author}
  {\bibfnamefont {A.}~\bibnamefont {S.~McLeod}}, \bibinfo {author}
  {\bibfnamefont {C.}~\bibnamefont {Wang}}, \bibinfo {author} {\bibfnamefont
  {M.~K.}\ \bibnamefont {Stewart}}, \bibinfo {author} {\bibfnamefont
  {Z.}~\bibnamefont {Zhao}}, \bibinfo {author} {\bibfnamefont {G.}~\bibnamefont
  {Dominguez}}, \bibinfo {author} {\bibfnamefont {M.}~\bibnamefont {Thiemens}},
  \bibinfo {author} {\bibfnamefont {M.~M.}\ \bibnamefont {Fogler}}, \bibinfo
  {author} {\bibfnamefont {M.~J.}\ \bibnamefont {Tauber}}, \bibinfo {author}
  {\bibfnamefont {A.~H.}\ \bibnamefont {Castro-Neto}}, \bibinfo {author}
  {\bibfnamefont {C.~N.}\ \bibnamefont {Lau}}, \bibinfo {author} {\bibfnamefont
  {F.}~\bibnamefont {Keilmann}}, \ and\ \bibinfo {author} {\bibfnamefont
  {D.~N.}\ \bibnamefont {Basov}},\ }\href {\doibase 10.1021/nl202362d}
  {\bibfield  {journal} {\bibinfo  {journal} {Nano Letters}\ }\textbf {\bibinfo
  {volume} {11}},\ \bibinfo {pages} {4701} (\bibinfo {year} {2011})},\ \bibinfo
  {note} {pMID: 21972938}\BibitemShut {NoStop}%
\bibitem [{\citenamefont {Fei}\ \emph {et~al.}(2012)\citenamefont {Fei},
  \citenamefont {Rodin}, \citenamefont {Andreev}, \citenamefont {Bao},
  \citenamefont {McLeod}, \citenamefont {Wagner}, \citenamefont {Zhang},
  \citenamefont {Zhao}, \citenamefont {Thiemens}, \citenamefont {Dominguez},
  \citenamefont {Fogler}, \citenamefont {Neto}, \citenamefont {Lau},
  \citenamefont {Keilmann},\ and\ \citenamefont {Basov}}]{10.1038/nature11253}%
  \BibitemOpen
  \bibfield  {author} {\bibinfo {author} {\bibfnamefont {Z.}~\bibnamefont
  {Fei}}, \bibinfo {author} {\bibfnamefont {A.~S.}\ \bibnamefont {Rodin}},
  \bibinfo {author} {\bibfnamefont {G.~O.}\ \bibnamefont {Andreev}}, \bibinfo
  {author} {\bibfnamefont {W.}~\bibnamefont {Bao}}, \bibinfo {author}
  {\bibfnamefont {A.~S.}\ \bibnamefont {McLeod}}, \bibinfo {author}
  {\bibfnamefont {M.}~\bibnamefont {Wagner}}, \bibinfo {author} {\bibfnamefont
  {L.~M.}\ \bibnamefont {Zhang}}, \bibinfo {author} {\bibfnamefont
  {Z.}~\bibnamefont {Zhao}}, \bibinfo {author} {\bibfnamefont {M.}~\bibnamefont
  {Thiemens}}, \bibinfo {author} {\bibfnamefont {G.}~\bibnamefont {Dominguez}},
  \bibinfo {author} {\bibfnamefont {M.~M.}\ \bibnamefont {Fogler}}, \bibinfo
  {author} {\bibfnamefont {A.~H.~C.}\ \bibnamefont {Neto}}, \bibinfo {author}
  {\bibfnamefont {C.~N.}\ \bibnamefont {Lau}}, \bibinfo {author} {\bibfnamefont
  {F.}~\bibnamefont {Keilmann}}, \ and\ \bibinfo {author} {\bibfnamefont
  {D.~N.}\ \bibnamefont {Basov}},\ }\href {\doibase 10.1038/nature11253}
  {\bibfield  {journal} {\bibinfo  {journal} {Nature}\ }\textbf {\bibinfo
  {volume} {487}},\ \bibinfo {pages} {82} (\bibinfo {year} {2012})}\BibitemShut
  {NoStop}%
\bibitem [{\citenamefont {Chen}\ \emph {et~al.}(2012)\citenamefont {Chen},
  \citenamefont {Badioli}, \citenamefont {Alonso-González}, \citenamefont
  {Thongrattanasiri}, \citenamefont {Huth}, \citenamefont {Osmond},
  \citenamefont {Spasenovi\'c}, \citenamefont {Centeno}, \citenamefont
  {Pesquera}, \citenamefont {Godignon}, \citenamefont {Elorza}, \citenamefont
  {Camara}, \citenamefont {de~Abajo}, \citenamefont {Hillenbrand},\ and\
  \citenamefont {Koppens}}]{10.1038/nature11254}%
  \BibitemOpen
  \bibfield  {author} {\bibinfo {author} {\bibfnamefont {J.}~\bibnamefont
  {Chen}}, \bibinfo {author} {\bibfnamefont {M.}~\bibnamefont {Badioli}},
  \bibinfo {author} {\bibfnamefont {P.}~\bibnamefont {Alonso-González}},
  \bibinfo {author} {\bibfnamefont {S.}~\bibnamefont {Thongrattanasiri}},
  \bibinfo {author} {\bibfnamefont {F.}~\bibnamefont {Huth}}, \bibinfo {author}
  {\bibfnamefont {J.}~\bibnamefont {Osmond}}, \bibinfo {author} {\bibfnamefont
  {M.}~\bibnamefont {Spasenovi\'c}}, \bibinfo {author} {\bibfnamefont
  {A.}~\bibnamefont {Centeno}}, \bibinfo {author} {\bibfnamefont
  {A.}~\bibnamefont {Pesquera}}, \bibinfo {author} {\bibfnamefont
  {P.}~\bibnamefont {Godignon}}, \bibinfo {author} {\bibfnamefont {A.~Z.}\
  \bibnamefont {Elorza}}, \bibinfo {author} {\bibfnamefont {N.}~\bibnamefont
  {Camara}}, \bibinfo {author} {\bibfnamefont {F.~J.~G.}\ \bibnamefont
  {de~Abajo}}, \bibinfo {author} {\bibfnamefont {R.}~\bibnamefont
  {Hillenbrand}}, \ and\ \bibinfo {author} {\bibfnamefont {F.~H.~L.}\
  \bibnamefont {Koppens}},\ }\href {\doibase 10.1038/nature11254} {\bibfield
  {journal} {\bibinfo  {journal} {Nature}\ }\textbf {\bibinfo {volume} {487}},\
  \bibinfo {pages} {77} (\bibinfo {year} {2012})}\BibitemShut {NoStop}%
\bibitem [{\citenamefont {Sounas}\ and\ \citenamefont
  {Caloz}(2011)}]{:/content/aip/journal/apl/99/23/10.1063/1.3665944}%
  \BibitemOpen
  \bibfield  {author} {\bibinfo {author} {\bibfnamefont {D.~L.}\ \bibnamefont
  {Sounas}}\ and\ \bibinfo {author} {\bibfnamefont {C.}~\bibnamefont {Caloz}},\
  }\href {\doibase http://dx.doi.org/10.1063/1.3665944} {\bibfield  {journal}
  {\bibinfo  {journal} {Applied Physics Letters}\ }\textbf {\bibinfo {volume}
  {99}},\ \bibinfo {eid} {231902} (\bibinfo {year} {2011})}\BibitemShut
  {NoStop}%
\bibitem [{\citenamefont {Fang}\ and\ \citenamefont {Sun}(2015)}]{lsa.2015.67}%
  \BibitemOpen
  \bibfield  {author} {\bibinfo {author} {\bibfnamefont {Y.}~\bibnamefont
  {Fang}}\ and\ \bibinfo {author} {\bibfnamefont {M.}~\bibnamefont {Sun}},\
  }\href {http://dx.doi.org/10.1038/lsa.2015.67} {\bibfield  {journal}
  {\bibinfo  {journal} {Light Sci Appl}\ }\textbf {\bibinfo {volume} {4}},\
  \bibinfo {pages} {e294} (\bibinfo {year} {2015})}\BibitemShut {NoStop}%
\bibitem [{\citenamefont {Nikitin}\ \emph {et~al.}(2011)\citenamefont
  {Nikitin}, \citenamefont {Guinea}, \citenamefont {Garc\'{i}a-Vidal},\ and\
  \citenamefont {Mart\'{i}n-Moreno}}]{PhysRevB.84.161407}%
  \BibitemOpen
  \bibfield  {author} {\bibinfo {author} {\bibfnamefont {A.~Y.}\ \bibnamefont
  {Nikitin}}, \bibinfo {author} {\bibfnamefont {F.}~\bibnamefont {Guinea}},
  \bibinfo {author} {\bibfnamefont {F.~J.}\ \bibnamefont {Garc\'{i}a-Vidal}}, \
  and\ \bibinfo {author} {\bibfnamefont {L.}~\bibnamefont
  {Mart\'{i}n-Moreno}},\ }\href {\doibase 10.1103/PhysRevB.84.161407}
  {\bibfield  {journal} {\bibinfo  {journal} {Phys. Rev. B}\ }\textbf {\bibinfo
  {volume} {84}},\ \bibinfo {pages} {161407} (\bibinfo {year}
  {2011})}\BibitemShut {NoStop}%
\bibitem [{\citenamefont {Christensen}\ \emph {et~al.}(2012)\citenamefont
  {Christensen}, \citenamefont {Manjavacas}, \citenamefont {Thongrattanasiri},
  \citenamefont {Koppens},\ and\ \citenamefont {García~de
  Abajo}}]{doi:10.1021/nn2037626}%
  \BibitemOpen
  \bibfield  {author} {\bibinfo {author} {\bibfnamefont {J.}~\bibnamefont
  {Christensen}}, \bibinfo {author} {\bibfnamefont {A.}~\bibnamefont
  {Manjavacas}}, \bibinfo {author} {\bibfnamefont {S.}~\bibnamefont
  {Thongrattanasiri}}, \bibinfo {author} {\bibfnamefont {F.~H.~L.}\
  \bibnamefont {Koppens}}, \ and\ \bibinfo {author} {\bibfnamefont {F.~J.}\
  \bibnamefont {García~de Abajo}},\ }\href {\doibase 10.1021/nn2037626}
  {\bibfield  {journal} {\bibinfo  {journal} {ACS Nano}\ }\textbf {\bibinfo
  {volume} {6}},\ \bibinfo {pages} {431} (\bibinfo {year} {2012})},\ \bibinfo
  {note} {pMID: 22147667}\BibitemShut {NoStop}%
\bibitem [{\citenamefont {Manjavacas}\ and\ \citenamefont {García~de
  Abajo}(2009)}]{doi:10.1021/nl802044t}%
  \BibitemOpen
  \bibfield  {author} {\bibinfo {author} {\bibfnamefont {A.}~\bibnamefont
  {Manjavacas}}\ and\ \bibinfo {author} {\bibfnamefont {F.~J.}\ \bibnamefont
  {García~de Abajo}},\ }\href {\doibase 10.1021/nl802044t} {\bibfield
  {journal} {\bibinfo  {journal} {Nano Letters}\ }\textbf {\bibinfo {volume}
  {9}},\ \bibinfo {pages} {1285} (\bibinfo {year} {2009})},\ \bibinfo {note}
  {pMID: 18672946}\BibitemShut {NoStop}%
\bibitem [{\citenamefont {Nikitin}\ \emph {et~al.}(2014)\citenamefont
  {Nikitin}, \citenamefont {Low},\ and\ \citenamefont
  {Martin-Moreno}}]{PhysRevB.90.041407}%
  \BibitemOpen
  \bibfield  {author} {\bibinfo {author} {\bibfnamefont {A.~Y.}\ \bibnamefont
  {Nikitin}}, \bibinfo {author} {\bibfnamefont {T.}~\bibnamefont {Low}}, \ and\
  \bibinfo {author} {\bibfnamefont {L.}~\bibnamefont {Martin-Moreno}},\ }\href
  {\doibase 10.1103/PhysRevB.90.041407} {\bibfield  {journal} {\bibinfo
  {journal} {Phys. Rev. B}\ }\textbf {\bibinfo {volume} {90}},\ \bibinfo
  {pages} {041407} (\bibinfo {year} {2014})}\BibitemShut {NoStop}%
\bibitem [{\citenamefont {Faraday}(1846)}]{Faraday}%
  \BibitemOpen
  \bibfield  {author} {\bibinfo {author} {\bibfnamefont {M.}~\bibnamefont
  {Faraday}},\ }\href@noop {} {\bibfield  {journal} {\bibinfo  {journal} {Phil.
  Trans. R. Soc.}\ }\textbf {\bibinfo {volume} {136}},\ \bibinfo {pages} {104}
  (\bibinfo {year} {1846})}\BibitemShut {NoStop}%
\bibitem [{\citenamefont {Suzuki}\ \emph {et~al.}(2003)\citenamefont {Suzuki},
  \citenamefont {ichi Fujii}, \citenamefont {Ohyama}, \citenamefont {Kobori},\
  and\ \citenamefont {Kotera}}]{doi:10.1143/JPSJ.72.3276}%
  \BibitemOpen
  \bibfield  {author} {\bibinfo {author} {\bibfnamefont {M.}~\bibnamefont
  {Suzuki}}, \bibinfo {author} {\bibfnamefont {K.}~\bibnamefont {ichi Fujii}},
  \bibinfo {author} {\bibfnamefont {T.}~\bibnamefont {Ohyama}}, \bibinfo
  {author} {\bibfnamefont {H.}~\bibnamefont {Kobori}}, \ and\ \bibinfo {author}
  {\bibfnamefont {N.}~\bibnamefont {Kotera}},\ }\href {\doibase
  10.1143/JPSJ.72.3276} {\bibfield  {journal} {\bibinfo  {journal} {Journal of
  the Physical Society of Japan}\ }\textbf {\bibinfo {volume} {72}},\ \bibinfo
  {pages} {3276} (\bibinfo {year} {2003})}\BibitemShut {NoStop}%
\bibitem [{\citenamefont {Crassee}\ \emph {et~al.}(2011)\citenamefont
  {Crassee}, \citenamefont {Levallois}, \citenamefont {Walter}, \citenamefont
  {Ostler}, \citenamefont {Bostwick}, \citenamefont {Rotenberg}, \citenamefont
  {Seyller}, \citenamefont {van~der Marel},\ and\ \citenamefont
  {Kuzmenko}}]{10.1038/nphys1816}%
  \BibitemOpen
  \bibfield  {author} {\bibinfo {author} {\bibfnamefont {I.}~\bibnamefont
  {Crassee}}, \bibinfo {author} {\bibfnamefont {J.}~\bibnamefont {Levallois}},
  \bibinfo {author} {\bibfnamefont {A.~L.}\ \bibnamefont {Walter}}, \bibinfo
  {author} {\bibfnamefont {M.}~\bibnamefont {Ostler}}, \bibinfo {author}
  {\bibfnamefont {A.}~\bibnamefont {Bostwick}}, \bibinfo {author}
  {\bibfnamefont {E.}~\bibnamefont {Rotenberg}}, \bibinfo {author}
  {\bibfnamefont {T.}~\bibnamefont {Seyller}}, \bibinfo {author} {\bibfnamefont
  {D.}~\bibnamefont {van~der Marel}}, \ and\ \bibinfo {author} {\bibfnamefont
  {A.~B.}\ \bibnamefont {Kuzmenko}},\ }\href {\doibase 10.1038/nphys1816}
  {\bibfield  {journal} {\bibinfo  {journal} {Nat. Phys.}\ }\textbf {\bibinfo
  {volume} {7}},\ \bibinfo {pages} {48} (\bibinfo {year} {2011})}\BibitemShut
  {NoStop}%
\bibitem [{\citenamefont {Ferreira}\ \emph {et~al.}(2011)\citenamefont
  {Ferreira}, \citenamefont {Viana-Gomes}, \citenamefont {Bludov},
  \citenamefont {Pereira}, \citenamefont {Peres},\ and\ \citenamefont
  {Castro~Neto}}]{PhysRevB.84.235410}%
  \BibitemOpen
  \bibfield  {author} {\bibinfo {author} {\bibfnamefont {A.}~\bibnamefont
  {Ferreira}}, \bibinfo {author} {\bibfnamefont {J.}~\bibnamefont
  {Viana-Gomes}}, \bibinfo {author} {\bibfnamefont {Y.~V.}\ \bibnamefont
  {Bludov}}, \bibinfo {author} {\bibfnamefont {V.}~\bibnamefont {Pereira}},
  \bibinfo {author} {\bibfnamefont {N.~M.~R.}\ \bibnamefont {Peres}}, \ and\
  \bibinfo {author} {\bibfnamefont {A.~H.}\ \bibnamefont {Castro~Neto}},\
  }\href {\doibase 10.1103/PhysRevB.84.235410} {\bibfield  {journal} {\bibinfo
  {journal} {Phys. Rev. B}\ }\textbf {\bibinfo {volume} {84}},\ \bibinfo
  {pages} {235410} (\bibinfo {year} {2011})}\BibitemShut {NoStop}%
\bibitem [{\citenamefont {Zhang}\ \emph {et~al.}(2012)\citenamefont {Zhang},
  \citenamefont {Zhou}, \citenamefont {Park}, \citenamefont {Rho},
  \citenamefont {Singh}, \citenamefont {Nam}, \citenamefont {Azad},
  \citenamefont {Chen}, \citenamefont {Yin}, \citenamefont {Taylor},\ and\
  \citenamefont {Zhang}}]{10.1038/ncomms1908}%
  \BibitemOpen
  \bibfield  {author} {\bibinfo {author} {\bibfnamefont {S.}~\bibnamefont
  {Zhang}}, \bibinfo {author} {\bibfnamefont {J.}~\bibnamefont {Zhou}},
  \bibinfo {author} {\bibfnamefont {Y.-S.}\ \bibnamefont {Park}}, \bibinfo
  {author} {\bibfnamefont {J.}~\bibnamefont {Rho}}, \bibinfo {author}
  {\bibfnamefont {R.}~\bibnamefont {Singh}}, \bibinfo {author} {\bibfnamefont
  {S.}~\bibnamefont {Nam}}, \bibinfo {author} {\bibfnamefont {A.~K.}\
  \bibnamefont {Azad}}, \bibinfo {author} {\bibfnamefont {H.-T.}\ \bibnamefont
  {Chen}}, \bibinfo {author} {\bibfnamefont {X.}~\bibnamefont {Yin}}, \bibinfo
  {author} {\bibfnamefont {A.~J.}\ \bibnamefont {Taylor}}, \ and\ \bibinfo
  {author} {\bibfnamefont {X.}~\bibnamefont {Zhang}},\ }\href
  {http://dx.doi.org/10.1038/ncomms1908} {\bibfield  {journal} {\bibinfo
  {journal} {Nat Commun}\ }\textbf {\bibinfo {volume} {3}},\ \bibinfo {pages}
  {942} (\bibinfo {year} {2012})}\BibitemShut {NoStop}%
\bibitem [{\citenamefont {Potton}(2004)}]{0034-4885-67-5-R03}%
  \BibitemOpen
  \bibfield  {author} {\bibinfo {author} {\bibfnamefont {R.~J.}\ \bibnamefont
  {Potton}},\ }\href {http://stacks.iop.org/0034-4885/67/i=5/a=R03} {\bibfield
  {journal} {\bibinfo  {journal} {Reports on Progress in Physics}\ }\textbf
  {\bibinfo {volume} {67}},\ \bibinfo {pages} {717} (\bibinfo {year}
  {2004})}\BibitemShut {NoStop}%
\bibitem [{\citenamefont {Sep\'ulveda}\ \emph {et~al.}(2010)\citenamefont
  {Sep\'ulveda}, \citenamefont {Gonz\'alez-D\'{i}az}, \citenamefont
  {Garc\'{i}a-Mart\'{i}n}, \citenamefont {Lechuga},\ and\ \citenamefont
  {Armelles}}]{PhysRevLett.104.147401}%
  \BibitemOpen
  \bibfield  {author} {\bibinfo {author} {\bibfnamefont {B.}~\bibnamefont
  {Sep\'ulveda}}, \bibinfo {author} {\bibfnamefont {J.~B.}\ \bibnamefont
  {Gonz\'alez-D\'{i}az}}, \bibinfo {author} {\bibfnamefont {A.}~\bibnamefont
  {Garc\'{i}a-Mart\'{i}n}}, \bibinfo {author} {\bibfnamefont {L.~M.}\
  \bibnamefont {Lechuga}}, \ and\ \bibinfo {author} {\bibfnamefont
  {G.}~\bibnamefont {Armelles}},\ }\href {\doibase
  10.1103/PhysRevLett.104.147401} {\bibfield  {journal} {\bibinfo  {journal}
  {Phys. Rev. Lett.}\ }\textbf {\bibinfo {volume} {104}},\ \bibinfo {pages}
  {147401} (\bibinfo {year} {2010})}\BibitemShut {NoStop}%
\bibitem [{\citenamefont {Hadad}\ \emph {et~al.}(2014)\citenamefont {Hadad},
  \citenamefont {Davoyan}, \citenamefont {Engheta},\ and\ \citenamefont
  {Steinberg}}]{doi:10.1021/ph500278w}%
  \BibitemOpen
  \bibfield  {author} {\bibinfo {author} {\bibfnamefont {Y.}~\bibnamefont
  {Hadad}}, \bibinfo {author} {\bibfnamefont {A.~R.}\ \bibnamefont {Davoyan}},
  \bibinfo {author} {\bibfnamefont {N.}~\bibnamefont {Engheta}}, \ and\
  \bibinfo {author} {\bibfnamefont {B.~Z.}\ \bibnamefont {Steinberg}},\ }\href
  {\doibase 10.1021/ph500278w} {\bibfield  {journal} {\bibinfo  {journal} {ACS
  Photonics}\ }\textbf {\bibinfo {volume} {1}},\ \bibinfo {pages} {1068}
  (\bibinfo {year} {2014})}\BibitemShut {NoStop}%
\bibitem [{\citenamefont {Tymchenko}\ \emph {et~al.}(2013)\citenamefont
  {Tymchenko}, \citenamefont {Nikitin},\ and\ \citenamefont
  {Martín-Moreno}}]{doi:10.1021/nn403282x}%
  \BibitemOpen
  \bibfield  {author} {\bibinfo {author} {\bibfnamefont {M.}~\bibnamefont
  {Tymchenko}}, \bibinfo {author} {\bibfnamefont {A.~Y.}\ \bibnamefont
  {Nikitin}}, \ and\ \bibinfo {author} {\bibfnamefont {L.}~\bibnamefont
  {Martín-Moreno}},\ }\href {\doibase 10.1021/nn403282x} {\bibfield  {journal}
  {\bibinfo  {journal} {ACS Nano}\ }\textbf {\bibinfo {volume} {7}},\ \bibinfo
  {pages} {9780} (\bibinfo {year} {2013})},\ \bibinfo {note} {pMID:
  24079266}\BibitemShut {NoStop}%
\bibitem [{\citenamefont {Maier}(2007)}]{MaierBook}%
  \BibitemOpen
  \bibfield  {author} {\bibinfo {author} {\bibfnamefont {S.~A.}\ \bibnamefont
  {Maier}},\ }\href@noop {} {\emph {\bibinfo {title} {Plasmonics: Fundamentals
  and Applications}}}\ (\bibinfo  {publisher} {Springer},\ \bibinfo {year}
  {2007})\BibitemShut {NoStop}%
\bibitem [{\citenamefont {Fallahi}\ and\ \citenamefont
  {Perruisseau-Carrier}(2012)}]{:/content/aip/journal/apl/101/23/10.1063/1.4769095}%
  \BibitemOpen
  \bibfield  {author} {\bibinfo {author} {\bibfnamefont {A.}~\bibnamefont
  {Fallahi}}\ and\ \bibinfo {author} {\bibfnamefont {J.}~\bibnamefont
  {Perruisseau-Carrier}},\ }\href {\doibase
  http://dx.doi.org/10.1063/1.4769095} {\bibfield  {journal} {\bibinfo
  {journal} {Applied Physics Letters}\ }\textbf {\bibinfo {volume} {101}},\
  \bibinfo {eid} {231605} (\bibinfo {year} {2012})}\BibitemShut {NoStop}%
\bibitem [{\citenamefont {Kildishev}\ \emph {et~al.}(2013)\citenamefont
  {Kildishev}, \citenamefont {Boltasseva},\ and\ \citenamefont
  {Shalaev}}]{Kildishev15032013}%
  \BibitemOpen
  \bibfield  {author} {\bibinfo {author} {\bibfnamefont {A.~V.}\ \bibnamefont
  {Kildishev}}, \bibinfo {author} {\bibfnamefont {A.}~\bibnamefont
  {Boltasseva}}, \ and\ \bibinfo {author} {\bibfnamefont {V.~M.}\ \bibnamefont
  {Shalaev}},\ }\href {\doibase 10.1126/science.1232009} {\bibfield  {journal}
  {\bibinfo  {journal} {Science}\ }\textbf {\bibinfo {volume} {339}} (\bibinfo
  {year} {2013}),\ 10.1126/science.1232009}\BibitemShut {NoStop}%
\bibitem [{\citenamefont {Song}\ and\ \citenamefont
  {Rudner}(2015)}]{arXiv:1506.04743}%
  \BibitemOpen
  \bibfield  {author} {\bibinfo {author} {\bibfnamefont {J.~C.~W.}\
  \bibnamefont {Song}}\ and\ \bibinfo {author} {\bibfnamefont {M.~S.}\
  \bibnamefont {Rudner}},\ }\href@noop {} {\bibfield  {journal} {\bibinfo
  {journal} {arXiv:1506.04743}\ } (\bibinfo {year} {2015})}\BibitemShut
  {NoStop}%
\end{thebibliography}%


\onecolumngrid
\clearpage
\begin{center}
\textbf{\large Supplemental Material}
\end{center}
\setcounter{equation}{0}
\setcounter{figure}{0}
\setcounter{table}{0}
\setcounter{page}{1}
\makeatletter
\renewcommand{\theequation}{S\arabic{equation}}
\renewcommand{\thefigure}{S\arabic{figure}}
\renewcommand{\bibnumfmt}[1]{[S#1]}
\renewcommand{\citenumfont}[1]{S#1}

\section{Von Neumann equation for density matrix}
Evolution of electron subsystem in massive Dirac system (MDS) in external electromagnetic field $\mathbf{E}$ can be described using the von Neumann equation for electron statistical operator
\begin{equation} \label{Eq:Liouville_eq}
i\hbar \frac{\partial \widehat{\rho}}{\partial t} =
[\mathcal{H}+V,\widehat{\rho}\,]
\end{equation}
where $\mathcal{H}$ is the single-electron Hamiltonian of MDS, $\widehat{V} = -e \mathbf{E}\cdot\mathbf{r}$ is the interaction term, and $e = -1.6\times 10^{-19}$ is the electron charge. In the basis of eigenfunctions of $\mathcal{H}$ Eq. \eqref{Eq:Liouville_eq} has the form
\begin{equation} \label{Eq:density_matrix_equation_en}
i\hbar \frac{\partial
	\rho_{\mathbf{k}j,\mathbf{k}'j'}}{\partial t} =
[\mathcal{H},\widehat{\rho}\,]_{\mathbf{k}j,\mathbf{k}'j'}+
[V,\widehat{\rho}\,]_{\mathbf{k}j,\mathbf{k}'j'},
\end{equation}
where 
\begin{equation} \label{Eq:Operator_matrix}
A_{\mathbf{k}j,\mathbf{k}'j'} = \int_S \Psi^*_{\mathbf{k}j}(\mathbf{r})\widehat{A}\Psi_{\mathbf{k}'j'}(\mathbf{r}) d^2 \mathbf{r},
\end{equation}
$\widehat{A} = \widehat{\rho}, \mathcal{H}, V$, $\mathbf{k}$ is two-dimensional electron wavevector, $j$ is an index designating quantum number of electrons in MDS. Integration is taken over the surface area $S$ of MDS. Functions $\Psi_{\mathbf{k}j}(\mathbf{r})$ are solutions of the equation 
\begin{equation} \label{Eq:Shrodinger_St_1}
\mathcal{H} \Psi_{\mathbf{k}j}(\mathbf{r}) = \mathcal{E}_{\mathbf{k}j}\Psi_{\mathbf{k}j}(\mathbf{r})
\end{equation} 
and can be written in the Bloch form
\begin{equation} \label{Eq:Graphene_solid_bloch}
\Psi_{\mathbf{k}j}(\mathbf{r})=\frac{1}{\sqrt{N}} \, e^{i
	\mathbf{k} \cdot\mathbf{r} } \, u_{\mathbf{k}j}(\mathbf{r}),
\end{equation}
where $N$ is the number of unit cells in MDS, and $u_{\mathbf{k}j}(\mathbf{r}) = u_{\mathbf{k}j}(\mathbf{r}+\mathbf{R}_n)$ is Bloch amplitude, $\mathbf{R}_n$ is the vector of MDS lattice. 

First term on right-hand side of Eq. \eqref{Eq:density_matrix_equation_en} can be written as
\begin{align}
[\mathcal{H},\widehat{\rho}\,]_{\mathbf{k}j,\mathbf{k}'j'} & = \rho_{\mathbf{k}j,\mathbf{k}'j'}
\left(\mathcal{E}_{\mathbf{k}j}-\mathcal{E}_{\mathbf{k}'j'}\right),
\end{align}
while second term takes form
\begin{equation} \label{Eq:Comut_inter}
[V,\widehat{\rho}\,]_{\mathbf{k}j,\mathbf{k}'j'} = \sum_{\mathbf{k}''j''} \left(V_{\mathbf{k}j,\mathbf{k}''j''}
\rho_{\mathbf{k}''j'',\mathbf{k}'j'} - \rho_{\mathbf{k}j,\mathbf{k}''j''} V_{\mathbf{k}''j'',\mathbf{k}'j'}\right).
\end{equation}
Let us consider first term on the left hand side of Eq. \eqref{Eq:Comut_inter},
\begin{align} 
& \sum_{\mathbf{k}''j''} V_{\mathbf{k}j,\mathbf{k}''j''} \rho_{\mathbf{k}''j'',\mathbf{k}'j'}  =  - e \mathbf{E}\cdot
\sum\limits_{\mathbf{k}''j''}\rho_{\mathbf{k}''j'',\mathbf{k}'j'} \int\limits_{S} \Psi_{\mathbf{k}j}^*(\mathbf{r}) \mathbf{r}\, \Psi_{\mathbf{k}''j''}(\mathbf{r}) d^2 \mathbf{r} \notag \\
&= \frac{i e}{\sqrt{N}}  \mathbf{E} \cdot\, \sum_{\mathbf{k}''j''} \rho_{\mathbf{k}''j'',\mathbf{k}'j'}
\int\limits_{S} \left[ - \frac{\partial u_{\mathbf{k}j}^*(\mathbf{r}) e^{- i \mathbf{k}\cdot\mathbf{r} }}{\partial \mathbf{k}}  \, \Psi_{\mathbf{k}''j''}(\mathbf{r}) + \frac{\partial u_{\mathbf{k}j}^*(\mathbf{r})}{\partial \mathbf{k}} e^{- i \mathbf{k}\cdot\mathbf{r} } \Psi_{\mathbf{k}''j''}(\mathbf{r}) \right] d^2 \mathbf{r}
\label{Eq_Ap:Promezh1} 
\end{align}
As electron wavefunctions are normalized, 
\begin{equation}
\int\limits_{S} \Psi_{\mathbf{k}j}^*(\mathbf{r}) \Psi_{\mathbf{k}'j'}(\mathbf{r})d^2 \mathbf{r} = \delta_{\mathbf{k} \mathbf{k}'} \delta_{j j'},
\end{equation}
the first term in square brackets in Eq. \eqref{Eq_Ap:Promezh1} takes form
\begin{equation} \label{Eq_Ap:Vspom1}
\frac{1}{\sqrt{N}}\sum_{\mathbf{k}''j''} \rho_{\mathbf{k}''j'',\mathbf{k}'j'}
\int\limits_{S} \frac{\partial u_{\mathbf{k}j}^*(\mathbf{r}) e^{- i \mathbf{k}\cdot\mathbf{r} }}{\partial \mathbf{k}}  \, \Psi_{\mathbf{k}''j''}(\mathbf{r}) d^2\mathbf{r} = \frac{\partial}{\partial \mathbf{k}}\sum_{\mathbf{k}''j''} \rho_{\mathbf{k}''j'',\mathbf{k}'j'}\int\limits_{S} \Psi_{\mathbf{k}j}^*(\mathbf{r}) \Psi_{\mathbf{k}''j''}(\mathbf{r}) d^2 \mathbf{r} =  \frac{\partial \rho_{\mathbf{k}j,\mathbf{k}'j'}}{\partial \mathbf{k}}.
\end{equation}
In order to simplify the second term in the square brackets we split integral over the surface area $S$ as sum of integrals over unit cells $\Omega_n$:
\begin{align}
&\frac{1}{N}\int\limits_{S} \frac{\partial 	u_{\mathbf{k}j}^*(\mathbf{r})}{\partial \mathbf{k}} u_{\mathbf{k}''j''}(\mathbf{r}) e^{i (\mathbf{k}''-	\mathbf{k} )\cdot\mathbf{r} } d^2 \mathbf{r} = \frac{1}{N} \sum_{n=1}^{N}\int\limits_{\Omega_n} \frac{\partial 	u_{\mathbf{k}j}^*(\mathbf{r})}{\partial \mathbf{k}} u_{\mathbf{k}''j''}(\mathbf{r}) e^{i (\mathbf{k}''-	\mathbf{k} )\cdot\mathbf{r} } d^2 \mathbf{r}  \notag \\
& = \frac{1}{N} \sum_{n=1}^{N}\int\limits_{\Omega_1} \frac{\partial u_{\mathbf{k}j}^*(\mathbf{r}+\mathbf{R}_n)}{\partial \mathbf{k}} u_{\mathbf{k}''j''}(\mathbf{r}+\mathbf{R}_n) e^{i 	(\mathbf{k}''-\mathbf{k})\cdot (\mathbf{r}+\mathbf{R}_n)} d^2 \mathbf{r}\notag \\
& = \frac{1}{N} \sum_{n=1}^{N} e^{i (\mathbf{k}'' - \mathbf{k})\cdot\mathbf{R}_n} \int\limits_{\Omega_1} \frac{\partial 	u_{\mathbf{k}j}^*(\mathbf{r})}{\partial \mathbf{k}} u_{\mathbf{k}''j''}(\mathbf{r}) e^{i (\mathbf{k}'' - \mathbf{k} ) \cdot \mathbf{r}} d^2 \mathbf{r} \notag \\
& = \delta_{\mathbf{k}\mathbf{k}''} \int_{\Omega_1} \frac{\partial	u_{\mathbf{k}j}^*(\mathbf{r})}{\partial \mathbf{k}}
u_{\mathbf{k}j''}(\mathbf{r})  d^2 \mathbf{r}= - \delta_{\mathbf{k}\mathbf{k}''} \int_{\Omega_1} u_{\mathbf{k}j}^*(\mathbf{r}) \frac{\partial
	u_{\mathbf{k}j''}(\mathbf{r})}{\partial \mathbf{k}} d^2 \mathbf{r}.\label{Eq_Ap:Vspom2}
\end{align}
In order to obtain last expression we integrated by parts and took into account periodicity of Bloch amplitudes at the unit cell boundaries. 

Plugging Eqs. \eqref{Eq_Ap:Vspom1}, \eqref{Eq_Ap:Vspom2} back into Eq. \eqref{Eq_Ap:Promezh1} we obtain
\begin{equation}
(V \widehat{\rho})_{\mathbf{k}j,\mathbf{k}'j'}=-i \,e \,\mathbf{E}\cdot\frac{\partial \rho_{\mathbf{k}j,\mathbf{k}'j'}}{\partial \mathbf{k}}-e \,\mathbf{E} \cdot \sum_{j''} \mathbf{R}_{j	j''}(\mathbf{k})\rho_{\mathbf{k}j'',\mathbf{k}'j'},
\label{interaction_comut_first}
\end{equation}
where 
\begin{equation}
\mathbf{R}_{j j''}(\mathbf{k}) = \frac{i }{2}\int_{\Omega_1} \left(u_{\mathbf{k}j}^*(\mathbf{r}) \frac{\partial	u_{\mathbf{k}j''}(\mathbf{r})}{\partial \mathbf{k}} -\frac{\partial u_{\mathbf{k}j}^*(\mathbf{r})}{\partial \mathbf{k}} u_{\mathbf{k}j''}(\mathbf{r})  \right) d^2 \mathbf{r}.
\label{r_definition}
\end{equation}
Repeating the steps above we also obtain
\begin{align}
(\widehat{\rho}\,V )_{\mathbf{k}j,\mathbf{k}'j'} & = i e \,\mathbf{E} \cdot \frac{\partial \rho_{\mathbf{k}j,\mathbf{k}'j'}}{\partial
	\mathbf{k}'}-e\, \mathbf{E} \cdot \sum_{j''} 	\rho_{\mathbf{k}j,\mathbf{k}'j''} \mathbf{R}_{j''j'}(\mathbf{k}'), \label{interaction_comut_second}
\end{align}
and 
\begin{align}
[V,\widehat{\rho}\,]_{\mathbf{k}j,\mathbf{k}'j'}&=-i \,e \,\mathbf{E}\cdot\left(\frac{\partial \rho_{\mathbf{k}j,\mathbf{k}'j'}}{\partial \mathbf{k}}+\frac{\partial\rho_{\mathbf{k}j,\mathbf{k}'j'}}{\partial	\mathbf{k}'}\right) -e \,\mathbf{E} \cdot\sum_{j''}\left(\mathbf{R}_{j j''}(\mathbf{k})\rho_{\mathbf{k}j'',\mathbf{k}'j'}-\rho_{\mathbf{k}j,\mathbf{k}'j''} \mathbf{R}_{j'' j'}(\mathbf{k}') \right) .
\label{interaction_comut_total}
\end{align}

Thus, the equation of motion for the electron density matrix takes form
\begin{align}
& i\hbar \frac{\partial \rho_{\mathbf{k}j,\mathbf{k}'j'}}{\partial t}=\rho_{\mathbf{k}j,\mathbf{k}'j'} 	\left(\mathcal{E}_{\mathbf{k}j}-\mathcal{E}_{\mathbf{k}'j'}\right)-i \,e \,\mathbf{E}\cdot\left(\frac{\partial\rho_{\mathbf{k}j,\mathbf{k}'j'}}{\partial \mathbf{k}}+ \frac{\partial \rho_{\mathbf{k}j,\mathbf{k}'j'}}{\partial	\mathbf{k}'}\right) \notag\\ 
&-e \,\mathbf{E} \cdot\sum_{j''}\left( \mathbf{R}_{j j''}(\mathbf{k})\rho_{\mathbf{k}j'',\mathbf{k}'j'}-\rho_{\mathbf{k}j,\mathbf{k}'j''}	\mathbf{R}_{j'' j'}(\mathbf{k}') \right).	\label{quantum_kinetic_spatial}
\end{align}
Eq. \eqref{quantum_kinetic_spatial} accounts both for the intraband motions and direct and indirect interband transitions. However, contribution of electrons indirect interband transitions to the optical response is negligible and thus can be omitted. Then, in the limit $\mathbf{k}'\to \mathbf{k}$, we obtain 
\begin{align}
\frac{\partial \rho_{jj'}(t,\mathbf{k})}{\partial t}+ \frac{e}{\hbar}\mathbf{E}\cdot\frac{\partial \rho_{jj'}(t,\mathbf{k})}{\partial \mathbf{k}} & = -\frac{i}{\hbar}\rho_{jj'}(t,\mathbf{k}) [\mathcal{E}_{\mathbf{k}j}-\mathcal{E}_{\mathbf{k}'j'}] \notag
\\& + \frac{i\,e}{\hbar} \,\mathbf{E} \cdot \sum_{j''}\left[ \mathbf{R}_{j j''}(\mathbf{k})\rho_{j''j'}(t,\mathbf{k})-\rho_{jj''}(t,\mathbf{k})
\mathbf{R}_{j'' j'}(\mathbf{k}) \right],
\label{quantum_kinetic_without}
\end{align}
where $\rho_{jj'}(t,\mathbf{k}) =  \rho_{\mathbf{k}j,\mathbf{k}'j'}(t)$, where $\nu,\mu = c,v$ denote electron and hole bands.  

\section{Incorporating relaxation into equations of motion for density matrix}
Electromagnetic radiation can not also couple electrons in different valleys in $\mathbf{k}$-space, and thus it is convenient to write equations separately for each of the valleys. We account for different valleys by introducing valley quantum number $\tau = \pm 1$ (or $K,K'$). For convenience, we designate density matrix elements as $ \rho_{jj'}(t,\mathbf{k}) = \rho^{\tau}_{\nu\mu}(t,\mathbf{k})$.

We introduce relaxation phenomenologically using relaxation time approximation. Then the equation for the density matrix of electrons in conduction band in $K$ valley takes form
\begin{align}
\label{Eq1}\frac{\partial \rho_{cc}^K(t,\mathbf{k})}{\partial t}+\frac{e}{\hbar}\mathbf{E}\cdot\frac{\partial \rho_{cc}^K (t,\mathbf{k})}{\partial \mathbf{k}}  = & -\frac{\rho_{cc}^K(t,\mathbf{k})-\rho_{cc}^{\mathrm{eq},K}(\mathbf{k})}{\tau_0}  + \frac{\rho_{cc}^{K'} (t,\mathbf{k}) - \rho_{cc}^K (t,\mathbf{k})}{\tau_1} \notag \\
&+ \frac{ i\,e}{\hbar} \,\mathbf{E} \cdot \left[ \mathbf{R}_{c v}^K(\mathbf{k})\rho_{vc}^K(t,\mathbf{k}) - \rho_{cv}^K(t,\mathbf{k}) \mathbf{R}_{v c}^K(\mathbf{k}) \right],
\end{align}
where $\rho_{cc}^{\mathrm{eq},K}(\mathbf{k})$ is equillibrium Fermi-Dirac distribution, $\tau_0$ is the population relaxation time. We also took into account that electron-phonon interactions can couple electrons in different valleys by introducing inter-valley scattering time $\tau_1$. The rest of equations for diagonal matrix elements can be written as
\begin{align}
\frac{\partial \rho_{cc}^{K'}}{\partial t}+\frac{e}{\hbar} \mathbf{E}\cdot\frac{\partial \rho_{cc}^{K'}}{\partial \mathbf{k}} & = -\frac{\rho_{cc}^{K'}-\rho_{cc}^{\mathrm{eq},K'}}{\tau_0} + \frac{\rho_{cc}^K - \rho_{cc}^{K'}}{\tau_1} + \frac{ i\,e}{\hbar} \,\mathbf{E} \cdot \left[ \mathbf{R}_{c v}^{K'}\rho_{vc}^{K'} - \rho_{cv}^{K'} \mathbf{R}_{v c}^{K'} \right], \\
\frac{\partial \rho_{vv}^K}{\partial t}+\frac{e}{\hbar} \mathbf{E}\cdot\frac{\partial \rho_{vv}^K}{\partial \mathbf{k}} & = -\frac{\rho_{vv}^K-\rho_{vv}^{\mathrm{eq},K}}{\tau_0} +\frac{\rho_{vv}^{K'} - \rho_{vv}^K}{\tau_1}  - \frac{i\,e}{\hbar} \,\mathbf{E} \cdot \left[ \mathbf{R}_{c v}^K\rho_{vc}^K-\rho_{cv}^K \mathbf{R}_{v c}^K \right], \\
\label{Eq4}\frac{\partial \rho_{vv}^{K'}}{\partial t}+\frac{e}{\hbar} \mathbf{E}\cdot\frac{\partial \rho_{vv}^{K'}}{\partial \mathbf{k}} & = -\frac{\rho_{vv}^{K'}-\rho_{vv}^{\mathrm{eq},{K'}}}{\tau_0} + \frac{\rho_{vv}^{K} - \rho_{vv}^{K'}}{\tau_1} - \frac{i\,e}{\hbar} \,\mathbf{E} \cdot \left[ \mathbf{R}_{c v}^{K'}\rho_{vc}^{K'}-\rho_{cv}^{K'} \mathbf{R}_{v c}^{K'} \right].
\end{align}
The non-diagonal matrix elements evolve as 
\begin{align}
\label{Eq5}\frac{\partial \rho_{cv}^K}{\partial t}+\frac{e}{\hbar} \mathbf{E}\cdot\frac{\partial \rho_{cv}^K}{\partial \mathbf{k}} & = -\left(i\omega_{cv}^K+\frac{1}{\tau_d}\right)\rho_{cv}^K  + \frac{2 i\,e}{\hbar} \,\mathbf{E}  \cdot \mathbf{R}_{c c}^K\rho_{cv}^K - \frac{i\,e}{\hbar} \,\mathbf{E} \cdot \mathbf{R}_{c v}^K \rho_{\mathrm{in}}^K, 
\end{align}
where $\rho_{\mathrm{in}}^K = \rho_{cc}^K - \rho_{vv}^K$, $\omega_{cv}^K = \left(\mathcal{E}_c^K - \mathcal{E}_v^K\right)/\hbar$, and $\tau_d$ is electron dephasing time. 

\section{Eigenenergies and eigenfunctions}
In order to calculate dipole matrix elements we need to know Bloch amplitudes $u_{\mathbf{k}j}$. We use MDS Hamiltonian 
\begin{align*}
{\cal H} = \frac{\Delta}{2}\sigma_z + t_0 a_0 \textbf{k}\cdot \boldsymbol{\sigma}_{\tau},
\end{align*}
where $\boldsymbol{\sigma}_{\tau}=(\tau\sigma_x,\sigma_y)$. After diagonalizing the Hamiltonian, we obtain eigenenergies
\begin{equation}
\mathcal{E}^{\tau}_{\mathbf{k}\nu} =\frac{\nu}{2}\sqrt{\beta^{\tau}_{\mathbf{k}}},
\end{equation}
and eigenfunctions
\begin{equation}
u^{\tau}_{\mathbf{k}c} = \frac{1}{\sqrt{S_{\Omega}}\,|\Phi^{\tau}_{\mathbf{k}}|} 
\left(
\begin{array}{c}
1 \\
W^{\tau}_{\mathbf{k}}
\end{array}
\right), \quad
u^{\tau}_{\mathbf{k}v} = \frac{1}{\sqrt{S_{\Omega}} \,|\Phi^{\tau}_{\mathbf{k}}|} 
\left(
\begin{array}{c}
-W^{\tau *}_{\mathbf{k}} \\
1
\end{array}
\right),
\end{equation}
where
\begin{align}
W^{\tau}_{\mathbf{k}} = \frac{2 t_0 a_0 \left(\tau k_x + i k_y \right) }{  \Delta  + \sqrt{\beta^{\tau}_{\mathbf{k}}}  }, \quad 
|\Phi^{\tau}_{\mathbf{k}}|^2 = 1 + |W^{\tau}_{\mathbf{k}}|^2,
\end{align}
\begin{equation}
\beta^{\tau}_{\mathbf{k}} = 4 a_0^2 t_0^2 k^2 + \Delta^2,
\end{equation}
$\nu=\pm 1$ denotes conduction and valence band respectively, and $S_{\Omega}$ is an area of unit cell.

\section{Dipole matrix elements $\mathbf{R}_{jj'}$}

\begin{align}
&\mathbf{R}_{cc}^{\tau}(\mathbf{k}) = \frac{i}{2} \int_{\Omega_1} \left(u_{\mathbf{k}c}^{\tau*} \frac{\partial u^{\tau}_{\mathbf{k}c}}{\partial \mathbf{k}} -\frac{\partial u_{\mathbf{k}c}^{\tau*}}{\partial \mathbf{k}} u_{\mathbf{k}c}^{\tau} \right) d^2 \mathbf{r} = \frac{i}{2|\Phi^{\tau}_{\mathbf{k}}|^2} \left(  W_{\mathbf{k}}^{\tau*} \frac{\partial W_{\mathbf{k}}^{\tau}}{\partial \mathbf{k}} -  W_{\mathbf{k}}^{\tau} \frac{\partial W_{\mathbf{k}}^{\tau*}}{\partial \mathbf{k}}\right) \notag \\
& =  - \frac{\tau \Omega_{\mathbf{k}\tau}^2 }{|\Phi^{\tau}_{\mathbf{k}}|^2}
\left(
\begin{array}{c}
-k_y \\
k_x
\end{array}
\right),
\end{align}

\begin{align}
&\mathbf{R}_{vv}^{\tau}(\mathbf{k}) = \frac{i}{2} \int_{\Omega_1} \left(u_{\mathbf{k}v}^{\tau*} \frac{\partial u^{\tau}_{\mathbf{k}v}}{\partial \mathbf{k}} -\frac{\partial u_{\mathbf{k}v}^{\tau*}}{\partial \mathbf{k}} u_{\mathbf{k}v}^{\tau} \right) d^2 \mathbf{r} = - \frac{i}{2|\Phi^{\tau}_{\mathbf{k}}|^2} \left(  W_{\mathbf{k}}^{\tau*} \frac{\partial W_{\mathbf{k}}^{\tau}}{\partial \mathbf{k}} -  W_{\mathbf{k}}^{\tau} \frac{\partial W_{\mathbf{k}}^{\tau*}}{\partial \mathbf{k}}\right) \notag \\
& =  \frac{\tau \Omega_{\mathbf{k}\tau}^2 }{|\Phi^{\tau}_{\mathbf{k}}|^2}
\left(
\begin{array}{c}
-k_y \\
k_x
\end{array}
\right),
\end{align}

\begin{align}
&\mathbf{R}_{cv}^{\tau}(\mathbf{k}) = \frac{i}{2} \int_{\Omega_1} \left(u_{\mathbf{k}c}^{\tau*} \frac{\partial u^{\tau}_{\mathbf{k}v}}{\partial \mathbf{k}} -\frac{\partial u_{\mathbf{k}c}^{\tau*}}{\partial \mathbf{k}} u_{\mathbf{k}v}^{\tau} \right) d^2 \mathbf{r} = - \frac{i}{|\Phi^{\tau}_{\mathbf{k}}|^2} \frac{\partial W_{\mathbf{k}}^{\tau*}}{\partial \mathbf{k}} \notag \\
& =  
 \frac{a_0 t }{\sqrt{\beta^{\tau}_{\mathbf{k}}} \alpha^{\tau}_{\mathbf{k}}}
\left(
\begin{array}{c}
4 a_0^2 t^2 k_x k_y - i\tau \left(\alpha^{\tau}_{\mathbf{k}} - 4 a_0^2 t^2 k_x^2\right)  \\
-\alpha^{\tau}_{\mathbf{k}} + 4 a_0^2 t^2 k_y^2 + 4 i \tau a_0^2 t^2 k_x k_y 
\end{array}
\right),
\end{align}
\begin{align}
\Omega_{\mathbf{k}\tau} = \frac{2 t_0 a_0 }{  \Delta + \sqrt{\beta^{\tau}_{\mathbf{k}}}  },
\end{align}
\begin{equation}
\alpha^{\tau}_{\mathbf{k}} = \sqrt{\beta^{\tau}_{\mathbf{k}}} \left(\sqrt{\beta^{\tau}_{\mathbf{k}}} + \Delta\right).
\end{equation}

Dot product $\mathbf{E}_0 \cdot \mathbf{R}_{c v}^{K,K'}$ defines valley interaction with a field of a given polarization. Let us assume that $\mathbf{E}_0 = |\mathbf{E}_0| (\mathbf{e}_x + i \mathbf{e}_y)$. Then for $K$ valley ($\tau_z = 1$) 
\begin{align}
\mathbf{E}_0 \cdot \mathbf{R}_{c v}^{K} =  |\mathbf{E}_0|  \frac{a_0 t \left( - 2 i \alpha^{K}_{\mathbf{k}} + 4 i a_0^2 t^2 (k_x^2 + k_y^2) \right)}{\sqrt{\beta^{K}_{\mathbf{k}}} \alpha^{K}_{\mathbf{k}}} 
\end{align}
In the case $k_x = k_y = 0$
\begin{align}
\mathbf{E}_0 \cdot \mathbf{R}_{c v}^{K} =  -2 i |\mathbf{E}_0| \frac{a_0 t}{\sqrt{\beta^{K}_{\mathbf{k}}} }.
\end{align}
Thus (+) polarized wave pumps K valley. Let us consider $K'$ valley ($\tau_z = - 1$)  now: 
\begin{align}
\mathbf{E}_0 \cdot \mathbf{R}_{c v}^{K'} =  |\mathbf{E}_0|  \frac{4 i a_0^3 t^3  (k_y^2  - k_x^2) }{\sqrt{\beta^{K}_{\mathbf{k}}} \alpha^{K}_{\mathbf{k}}}
\end{align}
In the case $k_x = k_y = 0$
\begin{align}
\mathbf{E}_0 \cdot \mathbf{R}_{c v}^{K'} = 0.
\end{align}
Thus electrons in K' valley do not interact with (+) polarized wave. 

\section{MDS interaction with laser pump}
Let us consider interaction of MDS with monochromatic electromagnetic wave
\begin{equation}
\mathbf{E} = \mathbf{E}_0 e^{-i \omega t} + c.c..
\end{equation}
We assume that frequency of the wave $\omega$ is in the vicinity of the frequency of electron interband transitions. Eqs. \eqref{Eq1}-\eqref{Eq5} are partial differential equations which have to be solved numerically in the general case. However, in the case of near-resonant wave interband electron transitions dominate intraband motion. Thus we can neglect terms containing partial derivatives over $\mathbf{k}$ on the right-hand side. 

Under the influence of electromagnetic field non-diagonal matrix elements evolve as $\rho_{cv} = \tilde{\rho}_{cv}(t) e^{-i\omega t}$, where $\tilde{\rho}_{cv}(t)$ is a slow varying amplitude. We are interested in the steady-state solution. Using the rotation-wave approximation and setting partial derivative of $\tilde{\rho}_{cv}(t)$ to zero, we obtain
\begin{align}
\tilde{\rho}_{cv}^{\tau} & =  \frac{e}{\hbar \left(\omega-\omega_{cv}^{\tau}+ i/\tau_d\right)} \,\mathbf{E}_0 \cdot \mathbf{R}_{c v}^{\tau} \rho_{\mathrm{in}}^{\tau}. \label{Eq:nondiag}
\end{align}
Eqs. \eqref{Eq1}-\eqref{Eq4} then take form
\begin{align}
\label{Eq:diag1} 0& = -\frac{\rho_{cc}^K-\rho_{cc}^{\mathrm{eq},K}}{\tau_0} + \frac{\rho_{cc}^{K'} - \rho_{cc}^K}{\tau_1}  + \frac{ i\,e}{\hbar} \, \left[ \mathbf{E}_0 \cdot \mathbf{R}_{c v}^K\tilde{\rho}_{cv}^{K*} - \mathbf{E}_0^*\cdot \mathbf{R}_{v c}^K  \tilde{\rho}_{cv}^K \right], \\
0 & = -\frac{\rho_{cc}^{K'}-\rho_{cc}^{\mathrm{eq},K'}}{\tau_0} +\frac{\rho_{cc}^K-\rho_{cc}^{K'}}{\tau_1} + \frac{ i\,e}{\hbar} \,\left[ \mathbf{E}_0 \cdot \mathbf{R}_{c v}^{K'}\tilde{\rho}_{cv}^{K'*} - \mathbf{E}_0^*\cdot \mathbf{R}_{v c}^{K'}  \tilde{\rho}_{cv}^{K'} \right], \\
0 & = -\frac{\rho_{vv}^K-\rho_{vv}^{\mathrm{eq},K}}{\tau_0} + \frac{\rho_{vv}^{K'}-\rho_{vv}^K}{\tau_1}  - \frac{i\,e}{\hbar} \, \left[ \mathbf{E}_0 \cdot \mathbf{R}_{c v}^K\tilde{\rho}_{cv}^{K*} - \mathbf{E}_0^*\cdot \mathbf{R}_{v c}^K  \tilde{\rho}_{cv}^K \right], \\
\label{Eq:diag4} 0 & = -\frac{\rho_{vv}^{K'}-\rho_{vv}^{\mathrm{eq},K'}}{\tau_0} +\frac{\rho_{vv}^{K} - \rho_{vv}^{K'}}{\tau_1}  - \frac{i\,e}{\hbar} \, \left[ \mathbf{E}_0 \cdot \mathbf{R}_{c v}^{K'}\tilde{\rho}_{cv}^{K'*} - \mathbf{E}_0^*\cdot \mathbf{R}_{v c}^{K'}  \tilde{\rho}_{cv}^{K'} \right].
\end{align}

By substituting Eq.~\eqref{Eq:nondiag} into equations \eqref{Eq:diag1}-\eqref{Eq:diag4}, we obtain system of four linear equations for diagonal matrix elements in two valleys
\begin{align*}
\rho_{vv}^K & =\beta_K \left(\rho_{vv}^{\mathrm{eq},K}  + \gamma \rho_{vv}^{K'} + \alpha_{K} \rho_{cc}^K\right), \\
\rho_{vv}^{K'} & = \beta_{K'}\left(\rho_{vv}^{\mathrm{eq},K'}  + \gamma \rho_{vv}^{K} + \alpha_{K'} \rho_{cc}^{K'}\right), \\
\rho_{cc}^K &  = \beta_K \left(\rho_{cc}^{\mathrm{eq},K} + \gamma \rho_{cc}^{K'} + \alpha_K \rho_{vv}^K \right), \\
\rho_{cc}^{K'} & = \beta_{K'} \left(\rho_{cc}^{\mathrm{eq},K'} + \gamma \rho_{cc}^{K} + \alpha_{K'} \rho_{vv}^{K'}\right)
\end{align*}
where 
\begin{align*}
\beta_K = \frac{1}{1 + \alpha_K + \gamma}, \qquad \gamma =  \frac{\tau_0}{\tau_1},
\end{align*}
\begin{equation*}
\alpha_{K,K'} = \frac{2 e^2 \tau_0 \tau_d |\mathbf{E}_0 \cdot \mathbf{R}_{c v}^{K,K'}|^2}{\hbar^2 \left( \tau_d^2(\omega-\omega_{cv}^{K,K'})^2 + 1\right)}.
\end{equation*}
In the paper we assumed that $\tau_d = \tau_0$.

\end{document}